\documentclass{emulateapj} \usepackage{apjfonts}

\newcommand{\be}{\begin{equation}}
\newcommand{\ee}{\end{equation}}

\def\ltsima{$\; \buildrel < \over \sim \;$}
\def\simlt{\lower.5ex\hbox{\ltsima}}
\def\gtsima{$\; \buildrel > \over \sim \;$}
\def\simgt{\lower.5ex\hbox{\gtsima}}

\begin{document}

\title{The speed of the `bullet' in the merging galaxy cluster 1E0657--56}

\author{Volker Springel\altaffilmark{1} \& Glennys
Farrar\altaffilmark{2}}

\altaffiltext{1} {Max-Planck-Institut f\"{u}r Astrophysik,
Karl-Schwarzschild-Strasse 1, 85740 Garching, Germany}

\altaffiltext{2} {Center for Cosmology and Particle Physics, New York
  University, New York, NY 10003}
 
\begin{abstract}
  Deep Chandra exposures of the hot galaxy cluster 1E0657--56 have
  revealed a highly dynamical state of the system due to an ongoing
  merger with a massive subcluster. The system is observed shortly
  after the first core-passage of the infalling subcluster, which
  moves approximately in the plane of the sky and is preceded by a
  prominent bow shock with Mach number ${\cal M}\sim 3$.  The inferred
  shock velocity of $\sim 4700\,{\rm km\,s^{-1}}$ has been commonly
  interpreted as the velocity of the `bullet' subcluster itself.  This
  velocity is unexpectedly high in the $\Lambda$CDM cosmology, which
  may require non-trivial modifications in the dark sector such as
  additional long-range scalar forces if taken at face value. Here we
  present explicit hydrodynamical toy models of galaxy cluster mergers
  which very well reproduce the observed dynamical state of 1E0657--56
  and the mass models inferred from gravitational lensing
  observations. However, despite a shock speed of $4500\,{\rm
  km\,s^{-1}}$, the subcluster's mass centroid is moving only with
  $\sim 2600\,{\rm km\,s^{-1}}$ in the rest frame of the system. The
  difference arises in part due to a gravitationally induced inflow
  velocity of the gas ahead of the shock towards the bullet, which
  amounts to $\sim 1100\,{\rm km\,s^{-1}}$ for our assumed $10:1$ mass
  ratio of the merger. A second effect is that the shock front moves
  faster than the subcluster itself, enlarging the distance between
  the subcluster and the bow shock with time. We also discuss the
  expected location of the lensing mass peak relative to the
  hydrodynamical features of the flow, and show that their spatial
  separation depends sensitively on the relative concentrations and
  gas fractions of the merging clusters, in addition to being highly
  time dependent. A generic $\Lambda$CDM collision model, where a
  bullet subcluster with concentration $c=7.2$ merges with a parent
  cluster with concentration $c=3$ on a zero-energy orbit, reproduces
  all the main observational features seen in 1E0657--56 with good
  accuracy, suggesting that 1E0657--56 is well in line with
  expectations from standard cosmological models. In theories with an
  additional ``fifth" force in the dark sector, the bullet subcluster
  can be accelerated beyond the velocity reached in $\Lambda$CDM, and
  the spatial offset between the X-ray peak and the mass centroid of
  the subcluster can be significantly enlarged. Our results stress the
  need for explicit hydrodynamical models for the proper
  interpretation of actively merging systems such as 1E0657--56.
\end{abstract}


\section{Introduction}
\label{intro}

The massive cluster of galaxies 1E0657--56 at $z=0.296$ offers a
so-far unique setting for testing many aspects of our understanding of
dark matter and baryonic physics.  In this system, a massive
sub-cluster (the "bullet" with $M_{200} \simeq 1.5 \times 10^{14}\,
M_{\sun}$) has fallen through the main cluster ($M_{200} \simeq 1.5
\times 10^{15}\, M_{\sun}$) on a trajectory nearly exactly in the
plane of the sky \citep{markevitch02,barrena02}.  The matter
distribution and masses quoted above have been determined from weak
and strong lensing studies \citep{clowe04,clowe06,bradac06}. Fitting
the large-field weak lensing data with two spherically symmetric NFW
mass distributions leads to (D. Clowe, private communication) $r_{200}
= 2136$ kpc, $c = 1.94$ and $r_{200} = 995$ kpc, $c = 7.12$, for the
main and subcluster respectively, taking $h_0 = 0.7$, $\Omega_{\rm m}
= 0.3$, $\Omega_{\Lambda} = 0.3$.  The distribution of baryons in the
diffuse intra-cluster medium (ICM) is obtained from $500\,{\rm ks}$
Chandra X-ray observations \citep{markevitch06}.  By fitting the
discontinuity in the gas properties across the prominent bow shock
preceding the gas bullet, the shock Mach number has been determined to
be ${\cal M}=3.0 \pm 0.4$ \citep{markevitch02,markevitch06}, which
together with a measurement of the pre-shock temperature translates to
an inferred shock velocity of $v_s = 4740^{+710}_{-550}\, {\rm km \,
s^{-1}}$ \citep{markevitch06}.  Note that the errors quoted in
\citep{markevitch06} were symmetrized and the central velocity rounded
down; the values used here are the most accurate ones (M. Markevitch,
private communication).

In previous work, it has been assumed that this shock velocity is
equal to the subcluster's relative velocity with respect to the parent
cluster \citep[][among
others]{markevitch02,markevitch06,hayashi06}. This velocity is much
higher than naturally expected according to the $\Lambda$CDM cosmology
\citep{hayashi06,farrar06}.  Using the large cosmological Millennium
simulation \citep{SpringelMill2005}, \citet{hayashi06} have measured
the distribution function of the velocity of the most massive dark
matter substructure in massive clusters of galaxies, fitting the
results with \be \log f(> V_{\rm sub}) = - \left(\frac{V_{\rm
sub}/V_{200}}{1.55}\right)^{3.3}, \label{eqnlikelihood} \ee where
$V_{\rm 200}$ is the circular velocity of the parent at the virial
radius, and $f(> V_{\rm sub})$ gives the fraction of clusters with a
most massive substructure moving faster than $V_{\rm sub}$.  Adopting
for $V_{\rm sub}$ a shock velocity of $4500\,{\rm km\,s^{-1}}$
\citep{markevitch04} and for the virial velocity of the parent cluster
$V_{200}= 2380\,{\rm km\,s^{-1}}$ (based on a mass model with
$M_{200}=2.16\times 10^{15}\,h^{-1}{M}_\odot$,
$R_{200}=1.64\,h^{-1}{\rm Mpc}$ and $h=0.73$), \citet{hayashi06} found
a relative likelihood of $f\simeq 0.01$ for 1E0657--56, suggesting
that the system, while rare, can be plausibly obtained in
$\Lambda$CDM.

However, as previously pointed out by \citet{farrar06} this estimate
appears overly optimistic in light of the new improved data on
1E0657--56. This data has increased the shock velocity to $4740\,{\rm
km\,s^{-1}}$, while the mass-estimate for $M_{\rm 200}$ has seen a
downward revision by a factor of $\sim 2$.  The virial velocity
$V_{200}$ that is normally associated with a mass $M_{200}$
\citep{Navarro1996} is given by $V_{200} = [10\, G\, H(z)\,
M_{200}]^{1/3}$.  Accounting for the revised lower mass estimates from
gravitational lensing \citep{clowe06,bradac06}, $V_{200}$ goes down to
at most $V_{200}\sim 1770\,{\rm km\,s^{-1}}$ at the cluster redshift
of $z=0.296$. This means that the critical ratio $V_{\rm sub}/V_{200}$
changes from 1.9 to 2.68, implying that the estimated likelihood of
seeing a subcluster as fast as the bullet in 1E0657--56 is really only
$f\simeq 0.8 \times 10^{-6}$, based on
Equation~(\ref{eqnlikelihood}). Given the low abundance of clusters as
massive as 1E0657--56, this is an exceedingly small probability which
cannot be straightforwardly understood in $\Lambda$CDM.

An attempt to estimate the relative velocity of the subcluster and
main cluster when separated by 720 kpc, taking reasonable initial
conditions and the measured mass distributions, finds a maximum
relative velocity of about $3500\,{\rm km\,s^{-1}}$ \citep{farrar06},
suggesting the possible presence of an additional long-range force
between dark matter concentrations if 4740 ${\rm km\,s^{-1}}$ is the
true relative velocity \citep[see][and references therein]{farrar06}.
The results of \citet{hayashi06} and the conclusion to be drawn from
\citet{farrar06} depend critically on the identification of the
velocity of the dark-matter bullet with the observationally inferred
shock velocity.  Given the importance of a reliable determination of
the bullet's velocity, we here explore the relation between the true
velocity of the dark matter subcluster and the properties of the shock
in the gas distribution, using simplified `toy' models of the
collision to make the physics clear.

We briefly summarize our methodology in \S\ref{meth}.  An analyses of the
hydrodynamical state of a best-matching merger simulation is given in
\S\ref{results}, which illustrates in particular that a significant difference
between the shock and bullet velocity is naturally expected.  In
\S\ref{SecMatch} we consider slight variations of our default model and
quantify in more detail the quality of the match to the observed 1E0657--56
system. This is followed with a discussion of the dependence on further model
parameters in \S\ref{SecParamDep}.  In \S\ref{fifthforce} we examine
simulations that include a fifth force between dark matter particles.
Finally, we conclude with a discussion in \S\ref{disc}.

\begin{figure*}[t]
\begin{center}
\resizebox{15.2cm}{!}{\includegraphics{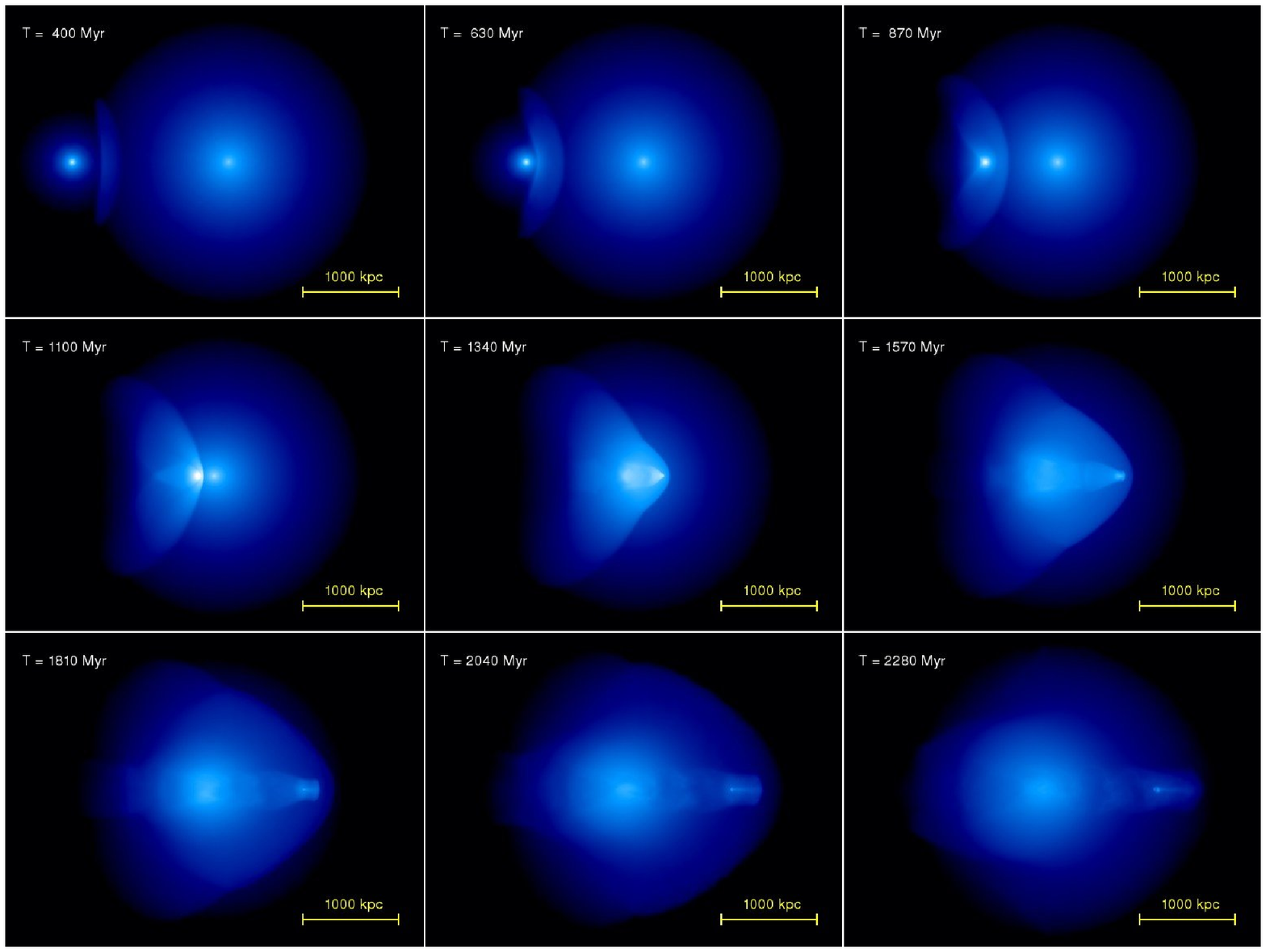}}\\
\ \\
\resizebox{15.2cm}{!}{\includegraphics{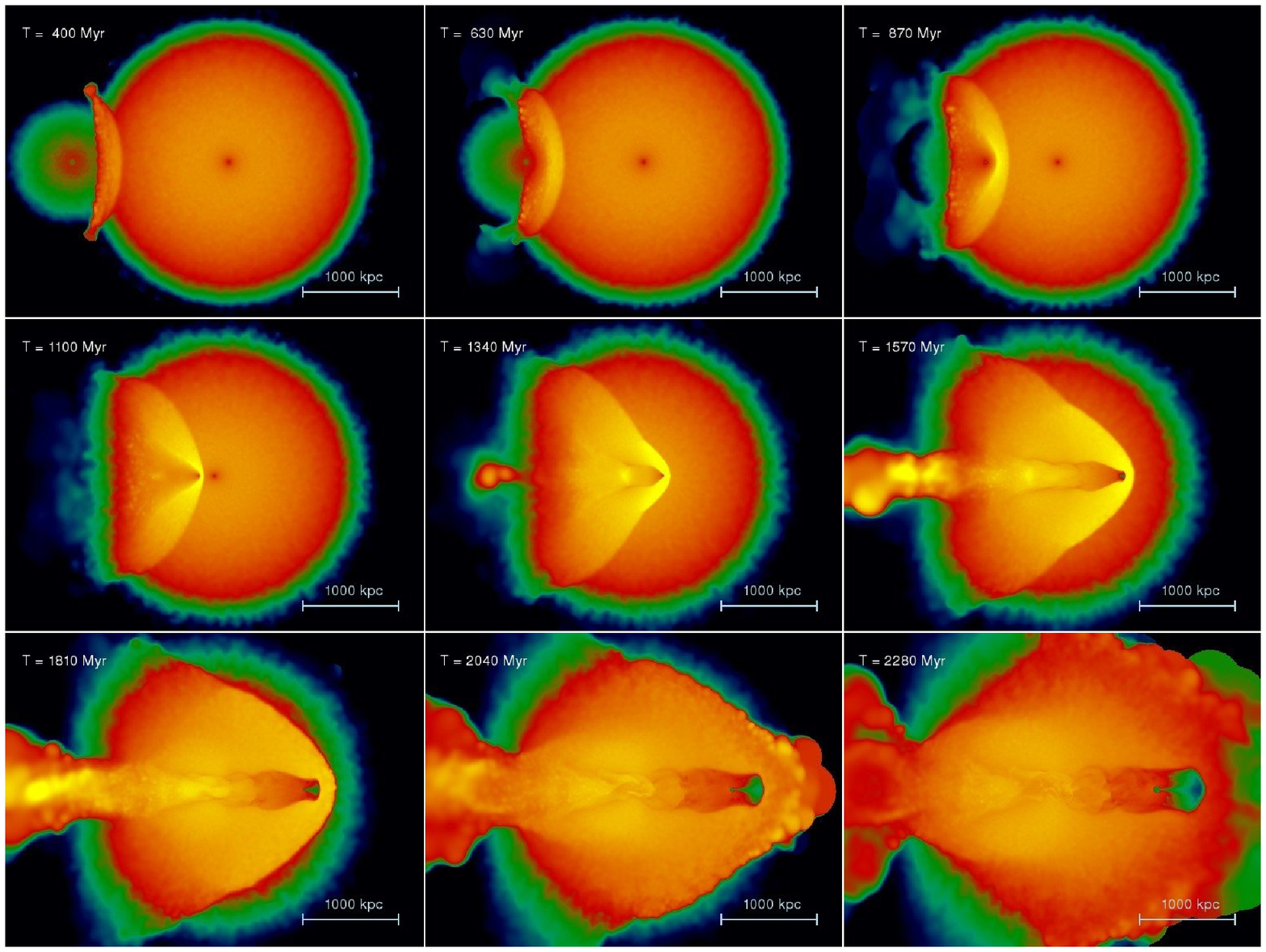}}
\end{center}
\caption{Time evolution of our default merger simulation, as seen in
  X-ray surface brightness (top) and in the luminosity-weighted
  projected temperature (bottom). The panel in the middle of both
  series of panels shows a time close to the best-match to the
  observed cluster 1E0657--56.}
\label{FigTimeEvolv}
\end{figure*}

\section{Numerical toy models for the bullet cluster} \label{meth}

\begin{figure*}[t]
\begin{center}
\resizebox{!}{6cm}{\includegraphics{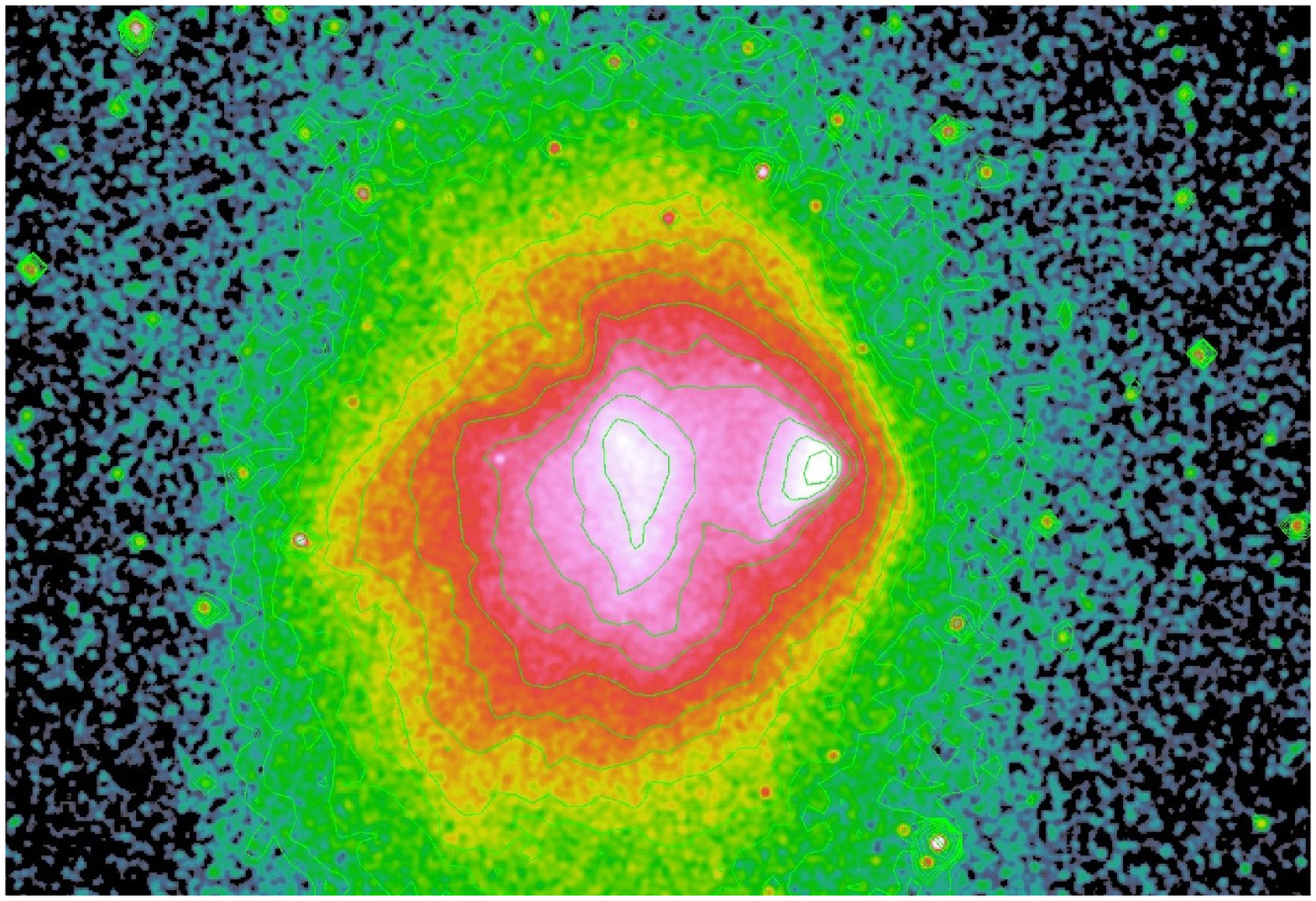}}\ %
\resizebox{!}{6cm}{\includegraphics{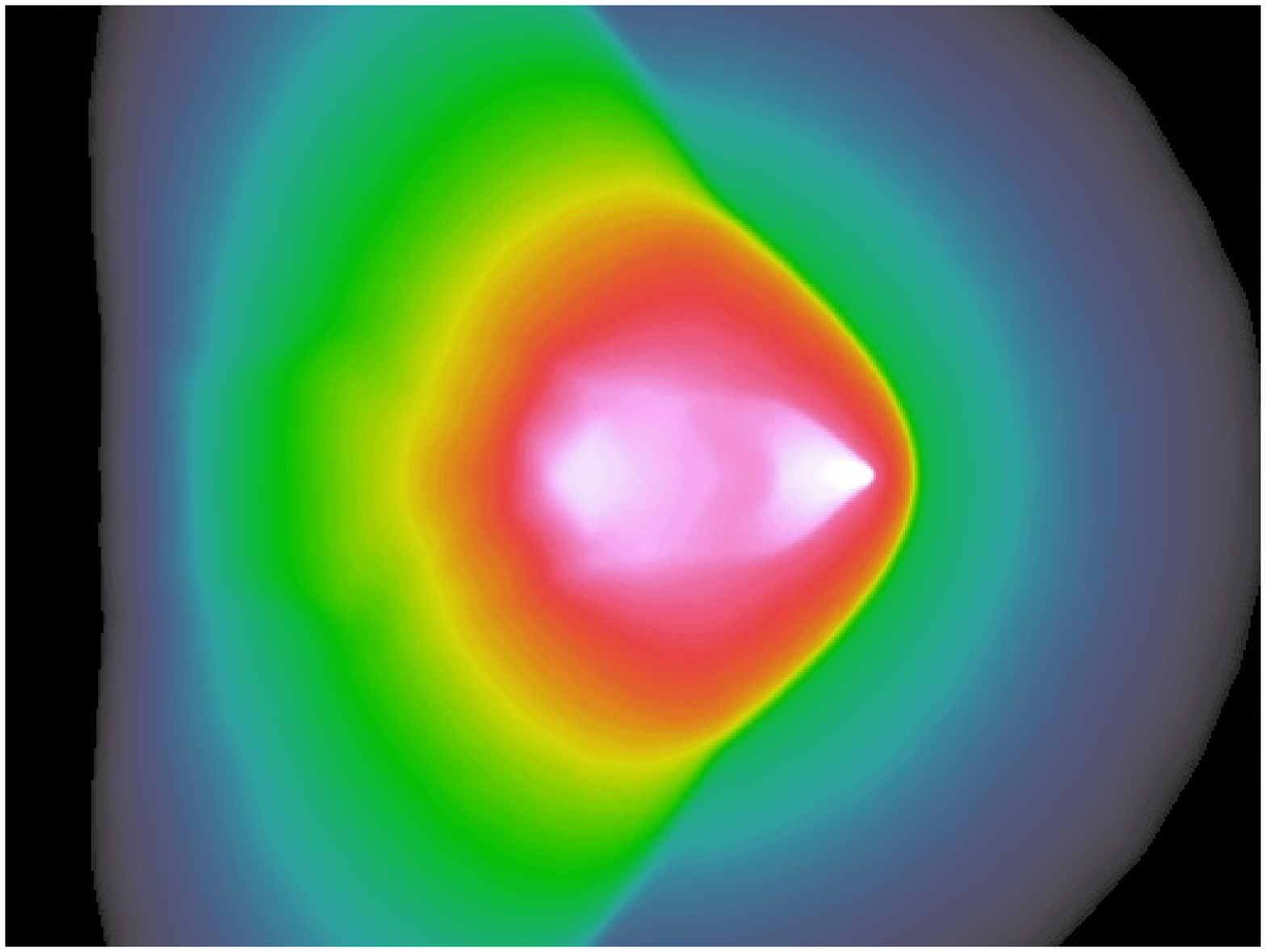}}\\
\end{center}
\caption{X-ray surface brightness of 1E0657--56 observed with Chandra (left).
  Clearly visible are two X-ray peaks. The wedge-like structure on the right
  associated with the `bullet' is bounded by a sharp contact discontinuity,
  which is interpreted as a cold front. The prominent bow shock in front
  demonstrates that the subcluster is moving to the right with high velocity.
  The panel on the right-hand side shows the X-ray surface brightness in one
  of our merger simulations, roughly drawn on the same scale and with similar
  dynamic range in the color table.}
\label{FigXrayMatch}
\end{figure*}

As our default model, we consider the collision of two isolated galaxy
clusters at $z=0$ of mass $M_{200}= 1.5 \times 10^{15}\, M_{\sun}$ and
$M_{200}= 1.5 \times 10^{14}\, M_{\sun}$, respectively, both
containing a universal baryon fraction of $f_{\rm b}=\Omega_{\rm
b}/\Omega_{\rm m} = 0.17$, consistent with the latest cosmological
constraints from the WMAP 3rd-year data \citep{spergel06}.  We take
the clusters to be spherically symmetric, with the dark matter
following an NFW-profile \citep{navarro97}, truncated at the virial
radius, and the gas being in hydrostatic equilibrium with a small core
in the center, but otherwise following the NFW profile as well.  For
the concentrations, we adopt the values $c=2.0$ and $c=7.2$ suggested
by the mass fits of Clowe (2006, private communication). Assuming a
conventional virial overdensity of 200 with respect to the critical
density, and a Hubble constant of $H_0=73\,{\rm km\,s^{-1}Mpc^{-1}}$,
the virial velocity and radius of the parent cluster are $1680\,{\rm
km\,s^{-1}}$ and $2300\,{\rm kpc}$, while those of the subcluster are
$780\,{\rm km\,s^{-1}}$ and $1070\,{\rm kpc}$, respectively.
Numerical realizations of these equilibrium clusters as N-body/SPH
models are constructed with the techniques described in
\cite{springel05}.

We assume that the clusters fall together from infinity on a
zero-energy orbit, giving the infalling subcluster a velocity of
$1870\,{\rm km\,s^{-1}}$ in the center-of-mass frame, at the time when
the virial radii first touch.  This is also the time when we start our
simulations. Note that all the absolute velocities we quote are
relative to the center-of-mass frame, hence the parent cluster moves
with $-187\,{\rm km\,s^{-1}}$ at the start. The simulations themselves
are evolved with a novel version of the cosmological smoothed particle
hydrodynamics (SPH) code {\small GADGET2}
\citep{springel01,springel05gadget}.  In our standard simulations, we
employ 4.4 million particles, equally divided between dark matter and
gas, and with equal mass resolution for the incoming subcluster and
the parent cluster. This results in a gas mass resolution of $m_{\rm
gas} = 1.275\times 10^{8}\, M_\sun$, and a dark matter resolution of
$m_{\rm dm}=6.225\times 10^{8}\, M_\sun$. We adopt a gravitational
softening length of $2.8\,{\rm kpc}$, which roughly demarks our
spatial resolution limit.

In our default calculation we consider a direct head-on encounter of
the two clusters, making the problem effectively axially symmetric
around the collision axis. The need to account for the non-linear
response of the dark matter during the merger requires
three-dimensional simulations nevertheless. We have also considered a
number of simulations with a small amount of orbital angular momentum
in the collision, measured by the minimum separation $b$ the
clusters would attain at perihelion if they were point masses. Also,
we considered variations of our initial conditions where the relative
concentrations and gas fractions of the initial cluster models were
altered, or where a fifth force between dark matter particles was
included.  Finally, for testing numerical convergence, we have also
carried out selected simulations with 35.2 million particles,
resulting in 8 times better mass resolution, and a 2 times better
spatial resolution in each dimension.

Since the measured temperature profile across the shock is consistent
with instant shock-heating of the electrons \citep{markevitch06}, we
model the ICM as an ideal gas with an adiabatic index of
$\gamma=5/3$. For simplicity, we neglect radiative cooling of the gas,
and consequently also star formation and the stellar systems of the
cluster galaxies. We also disregard magnetic fields, which could in
principle play an important role in stabilizing cold front against
fluid instabilities and thermal conduction
\citep{Ettori2000,Vikhlinin2001b,Lyutikov2006}. Our simulations are
carried out in physical coordinates, ignoring the cosmological
background expansion, the infall of other structures, and the
accretion of diffuse matter. While our approach is much more
restricted than fully self-consistent cosmological modeling, the short
time required for completing the merger (of the order of a few Gyrs)
should make it a reasonably good approximation.  At the same time, the
simplicity of the toy model makes the physical phenomena easier to
identify, in particular with respect to the relation between observed
shock properties and the underlying kinematics of the dark matter,
which is our main interest here.  We shall begin with the simplest of
the toy models, and then explore how the conclusions change as the
model is varied.

\begin{figure*}[t]
\begin{center}
\vspace*{-1.3cm}\resizebox{17.4cm}{!}{\includegraphics{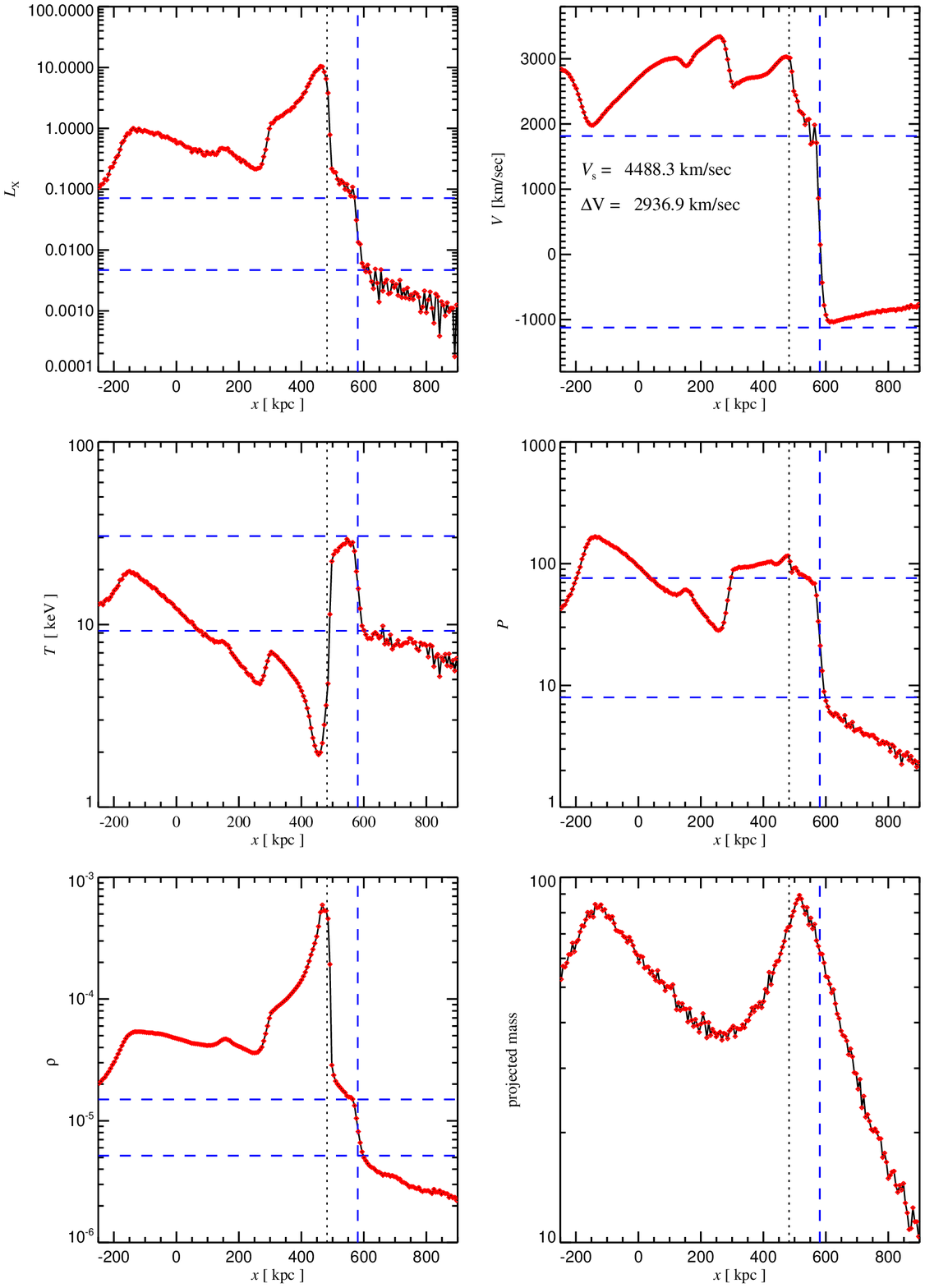}}\\
\end{center}
\caption{Local profiles of different hydrodynamical quantities along the
  symmetry axis through the nose of the shock, at time $t=1340\,{\rm Myr}$ after start of
  the simulation, corresponding to 180 Myr after core passage.  From top to
  bottom, we show the X-ray emissivity, gas streaming velocity along the
  $x$-direction, gas temperature, gas pressure, and gas density. Finally, the
  bottom right panel shows the projected total mass surface density along the
  $x$-direction. The dashed vertical line indicates the location of 
  the bow shock, while the horizontal dashed lines give the expected jump in the
  respective quantity for a Mach number of ${\cal M}=2.8$. The vertical dotted line
  marks the location of the edge of the bullet.}
\label{FigProfiles}
\end{figure*}

\begin{figure}[t]
\begin{center}
\resizebox{8.5cm}{!}{\includegraphics{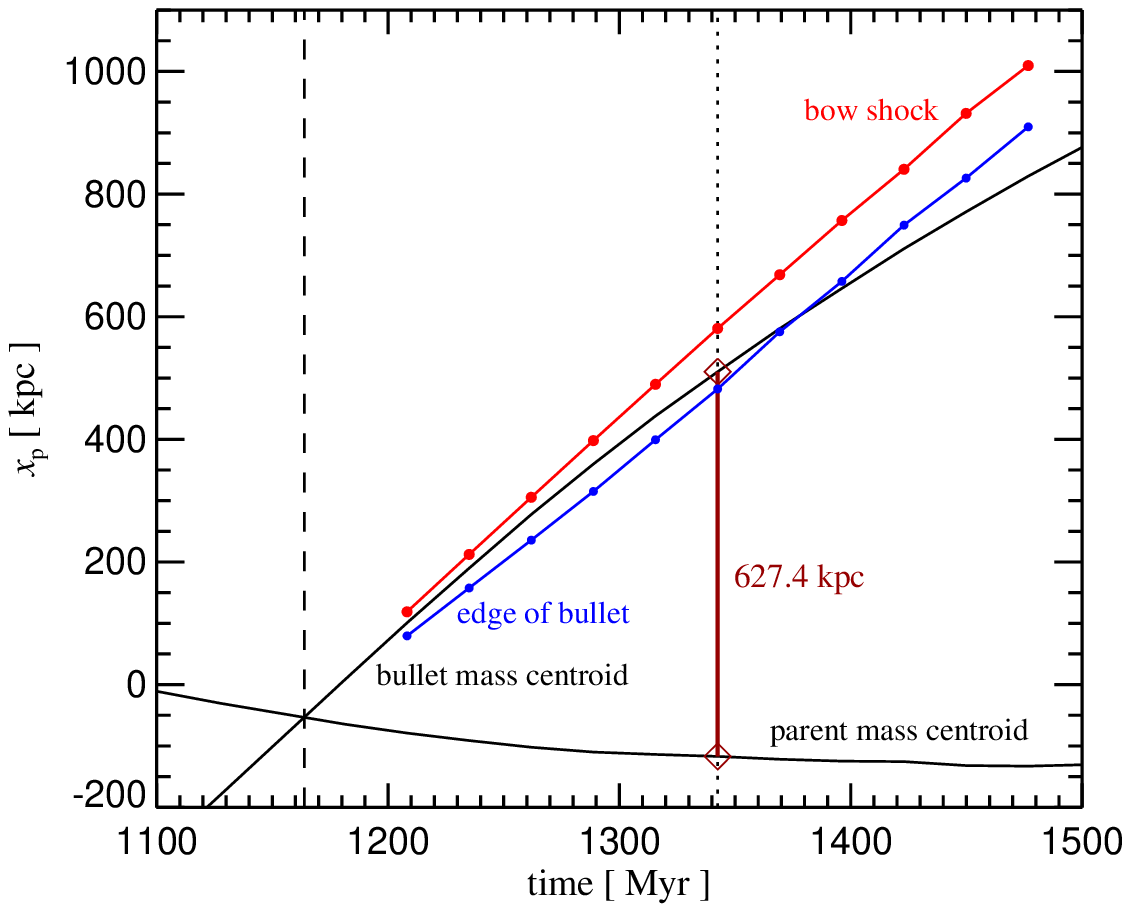}}\\
\resizebox{8.5cm}{!}{\includegraphics{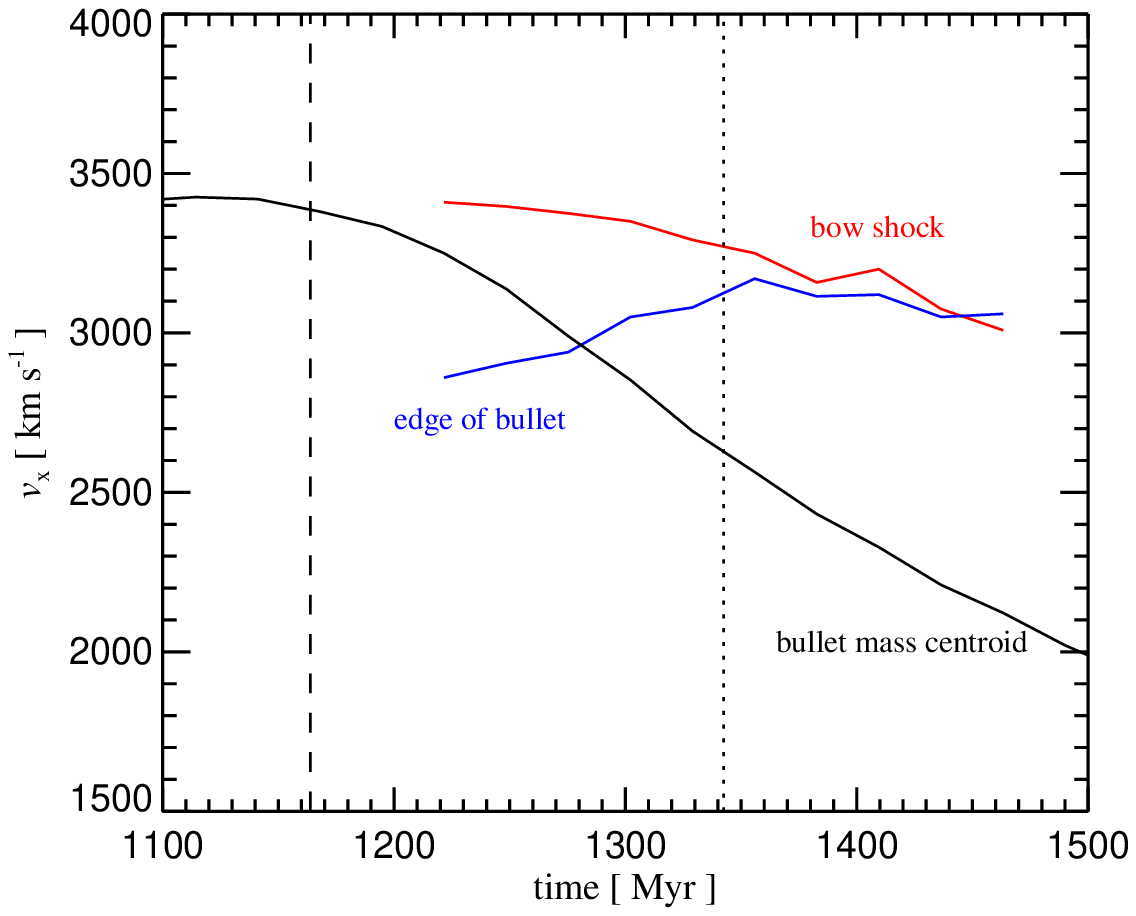}}\\
\end{center}
\caption{The top panel shows the location of the bow shock (red) and the edge
  of the bullet (blue) as a function of time. Also shown are the positions of
  the mass centroids of the bullet and the cluster with solid lines (as
  labeled).  In the bottom, the corresponding velocities are shown, obtained
  by numerical differentiation of the measured locations as a function of
  time. The time of core passage is marked with a dashed vertical line, while
  the time of the best fit of our simulation to the observed bullet cluster is
  marked with a dotted line.}
\label{FigSeparations}
\end{figure}

\section{Results for the basic toy model}  \label{results}

In Figure~\ref{FigTimeEvolv}, we show the time evolution of our
default simulation model in the form of maps of X-ray surface
brightness and projected luminosity-weighted temperature\footnote{A
video of the merger simulation is available for download at {\tt
http://www.mpa-garching.mpg.de/$\sim$volker/bullet}}.  We only
consider simple bremsstrahlung, with an X-ray volume emissivity
$\epsilon_X\propto \rho_{\rm g}^2\sqrt{T}$, where $\rho_{\rm g}$ and
$T$ are the local gas density and temperature, respectively. The
bullet cluster enters the parent at time $t =0$ from the left, and
passes the core at around $t \sim 1200\, {\rm Myr}$.  The bullet is
preceeded by a prominent bow shock, which develops a `bend' during
core passage, when the central region of the shock front speeds
up. The compression of the central gas during core passage makes the
centers of the parent and the bullet cluster very bright in X-rays for
a few hundred Myrs. In addition, the bullet develops a prominent 'cold
front' \citep{Vikhlinin2001} after core passage, which can be seen as
a sharp wedge-like `edge' in the X-ray surface brightness. The
temperature maps show the strong heating of the gas by the shock
front, while the bullet itself is seen to be quite cold, revealing
that the wedge-like edge of the bullet is really a cold front
\citep[see][for a review]{Markevitch2007}. Note that the maps shown in
Fig.~\ref{FigTimeEvolv} extend to substantially lower surface
brightness levels than are observationally accessible.

In Figure~\ref{FigXrayMatch}, we show an observed X-ray map of
1E0657--56 \citep[based on data presented in][]{markevitch06} and
compare it to a map of the simulation drawn with a dynamic range that
roughly matches that accessible in the observations. We selected a
time that provides approximately the best match to the observed X-ray
morphology, as will be discussed in more detail later. The simulation
clearly shows a double peaked structure in the X-ray emission similar
to that observed, with a wedge-like shape for the bullet that moves to
the right, and with a prominent bow shock in front of the bullet. This
broad agreement reconfirms the interpretation of 1E0657--56 as a
merging system.  We now turn to a more detailed analysis of the
simulation results, in order to evaluate how well they can
quantitatively reproduce the properties of the observed system.

\subsection{The dynamical structure of the merging system}

A much clearer understanding of the hydrodynamics involved is obtained by
looking at profiles of the gas and the dark matter distribution along a line
through the nose of the shock. This is done in Figure~\ref{FigProfiles} for
the local X-ray emissivity, gas velocity field, temperature, pressure, gas
density, and total projected mass.  In this figure at time $t=1340\,{\rm Myr}$
(corresponding to the central panel in the time evolution of Fig.~1), we can
clearly identify the location of the bow shock at $x_s\simeq 580\,{\rm kpc}$,
and a contact discontinuity at $x_c\simeq 482\,{\rm kpc}$.  The latter marks
the edge of the bullet; behind it, the gas is quite dense and cold, and very
bright in X-rays.  The pressure is essentially continuous across this front.

At the bow shock, the density, velocity, pressure, temperature, and
X-ray emissivity are all discontinuous (as is the specific
entropy). The sizes of the corresponding jumps are consistently
described by the Rankine-Hugoniot relations for an adopted Mach number
of ${\cal M}=2.8$, as indicated in the panels of the figure. The
preshock-temperature is $\simeq 9\,{\rm kev}$, just what is observed,
corresponding to a pre-shock soundspeed of $c_s = 1600\,{\rm
km\,s^{-1}}$. The inferred shock velocity is hence $v_s \equiv {\cal
M}\,c_s = 4488\,{\rm km\,s^{-1}}$. Note that a slightly higher Mach
number of 3.0 matches the numerical results still reasonably well, and
would then in fact result in a shock velocity exactly equal to the
value inferred from the most recent observational estimates
(references).  In any case, the value of ${\cal M}=2.8$ is well inside
the observational uncertainty.

\begin{figure*}[t]
\begin{center}
\resizebox{5.6cm}{!}{\includegraphics{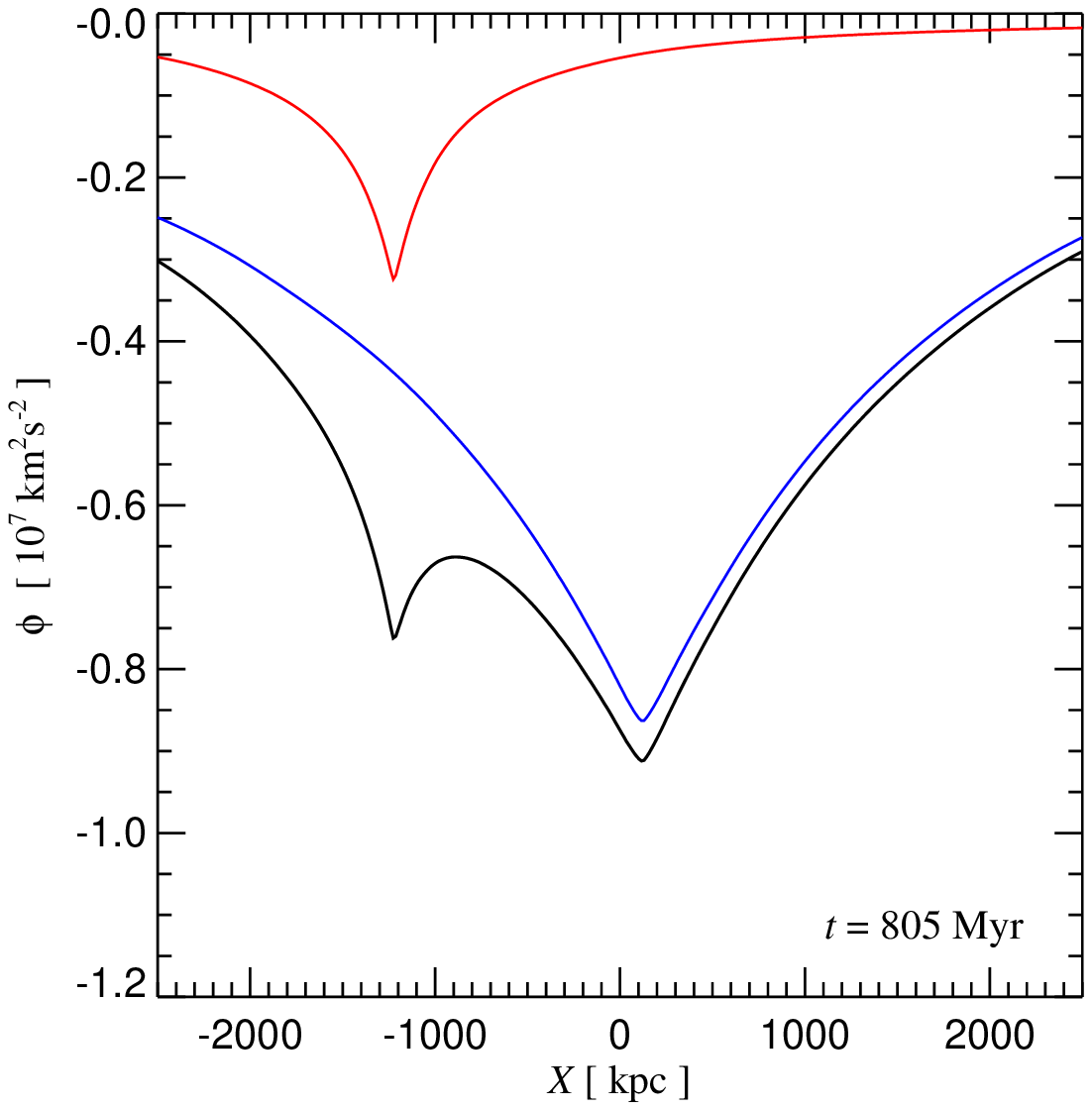}}\ %
\resizebox{5.6cm}{!}{\includegraphics{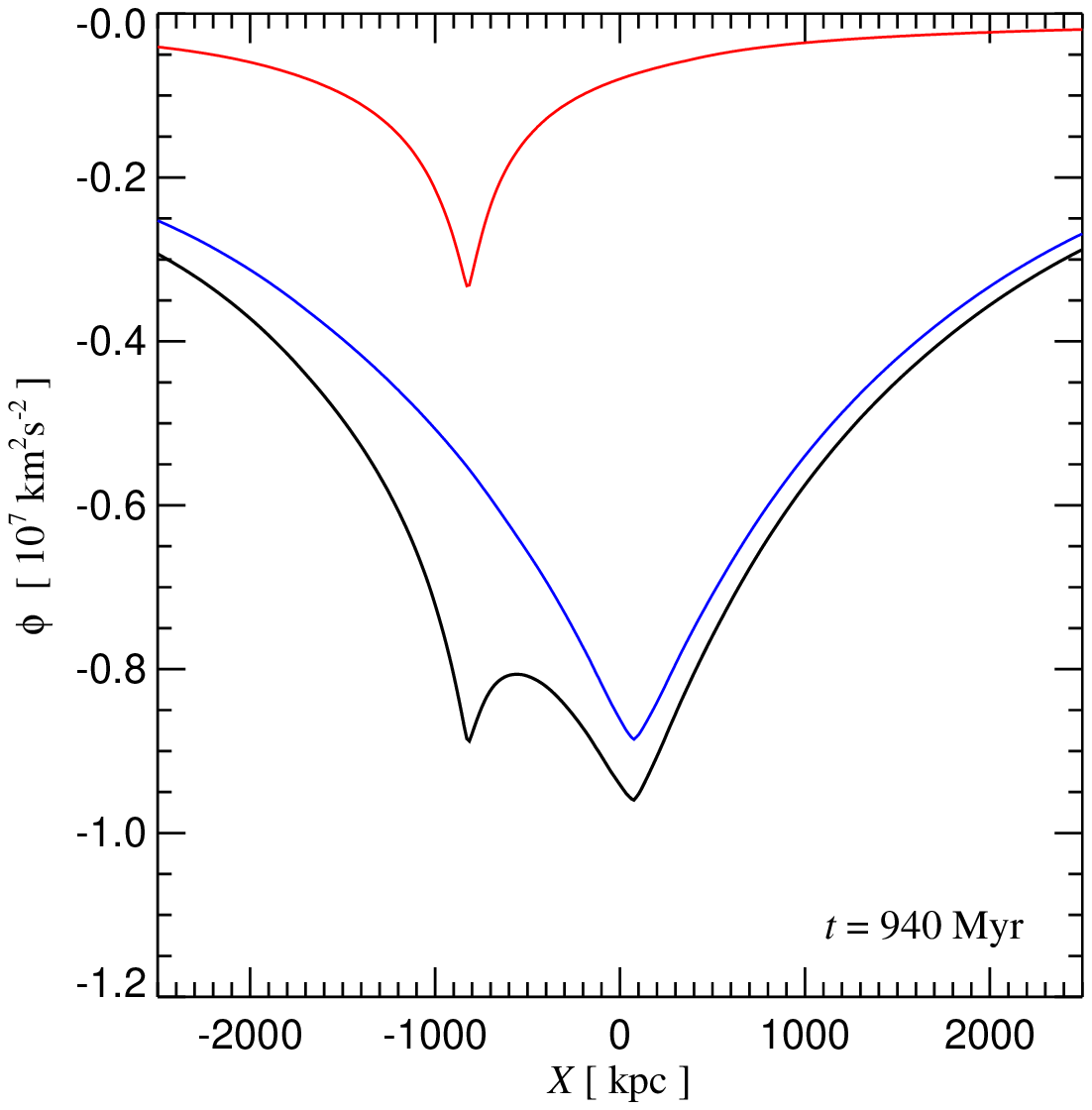}}\ %
\resizebox{5.6cm}{!}{\includegraphics{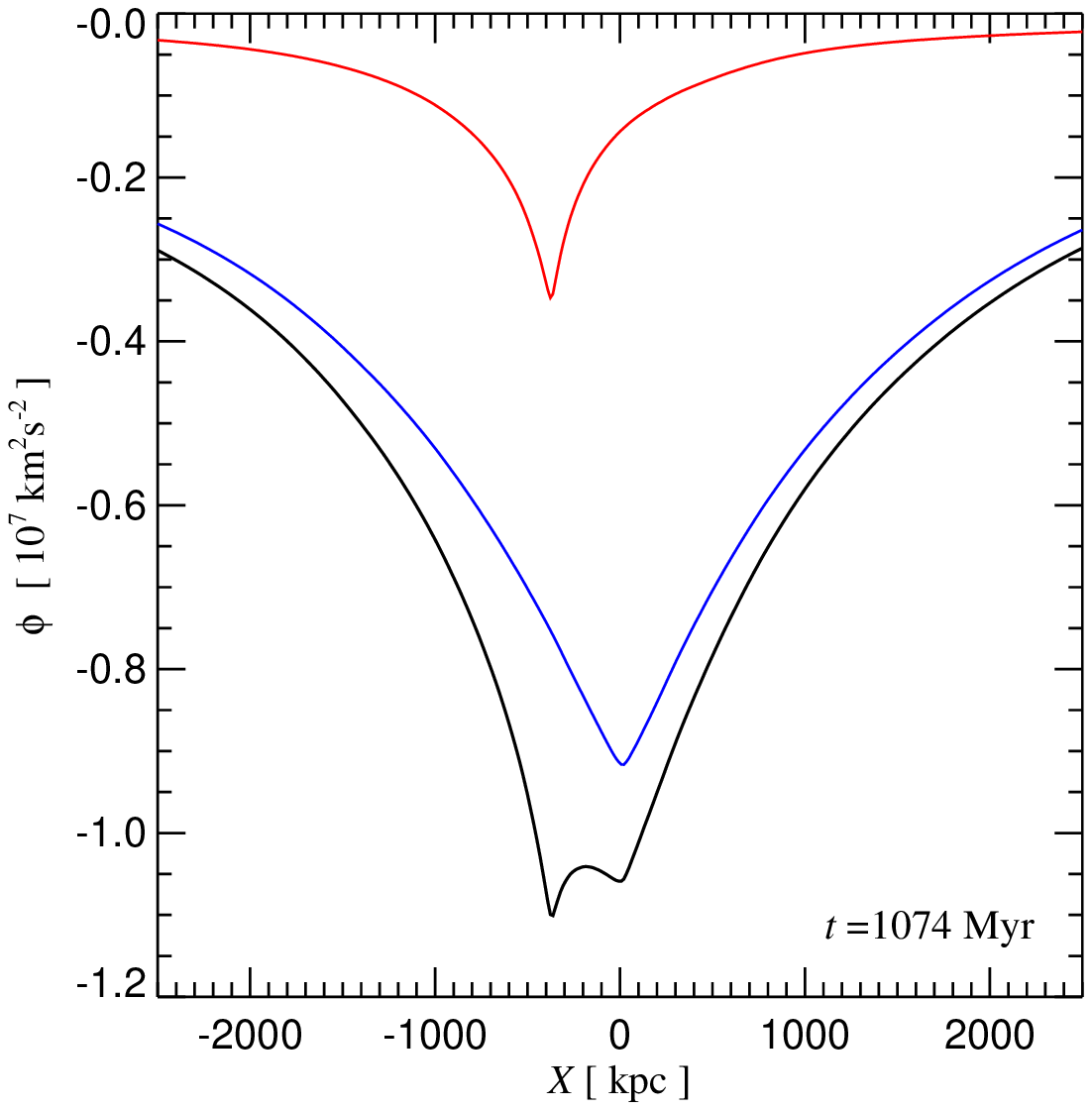}}\\%
\resizebox{5.6cm}{!}{\includegraphics{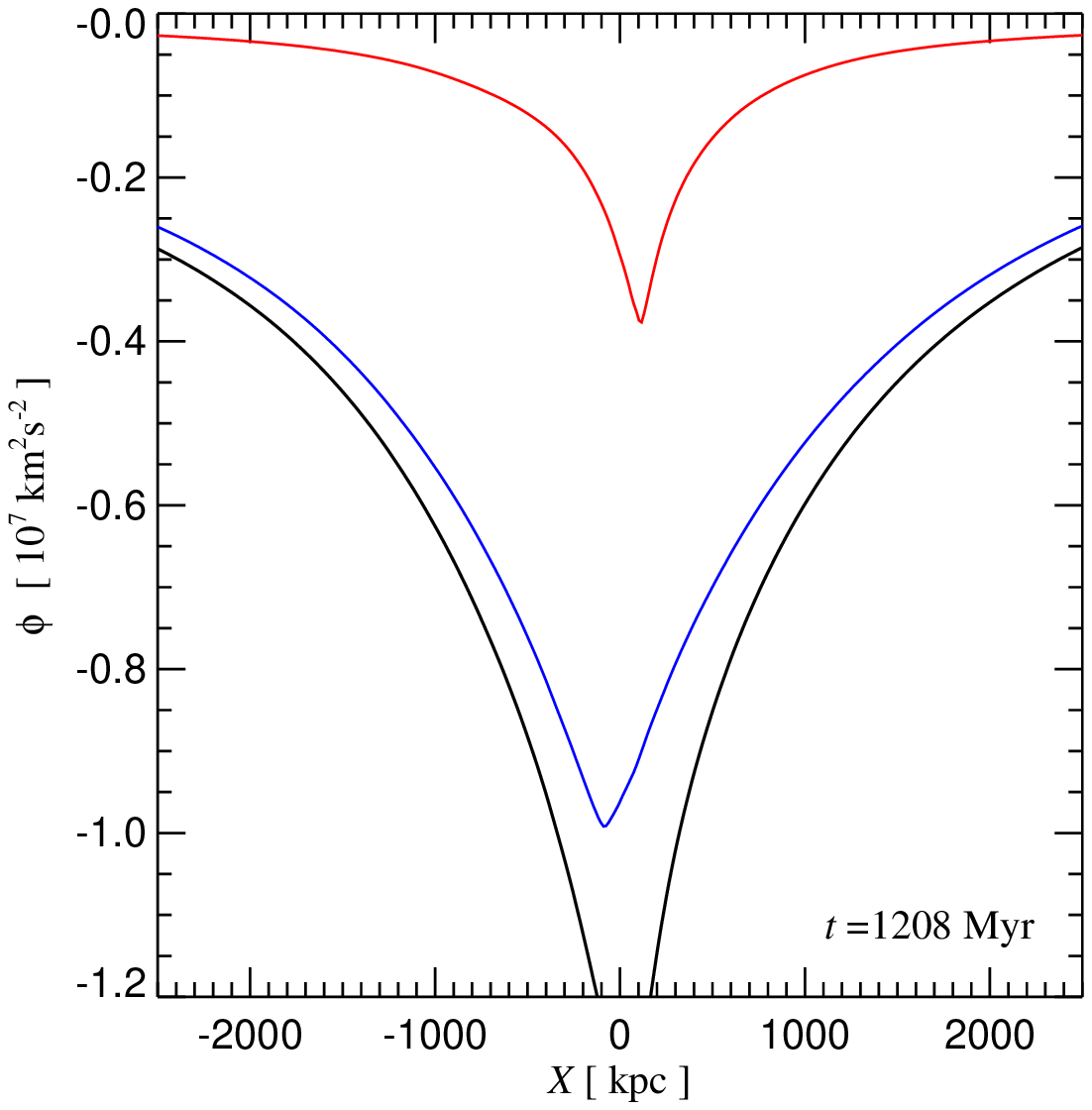}}\ %
\resizebox{5.6cm}{!}{\includegraphics{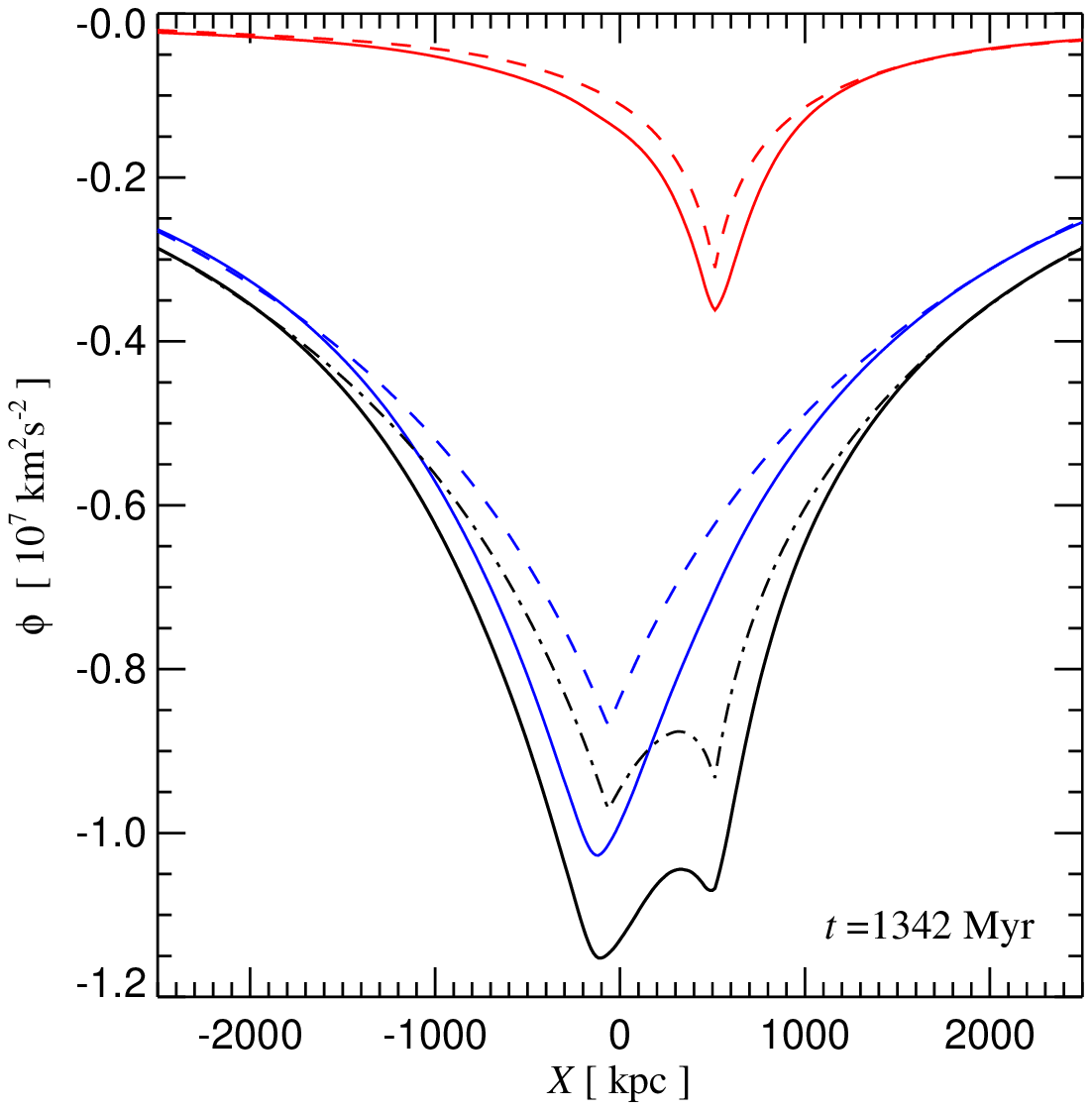}}\ %
\resizebox{5.6cm}{!}{\includegraphics{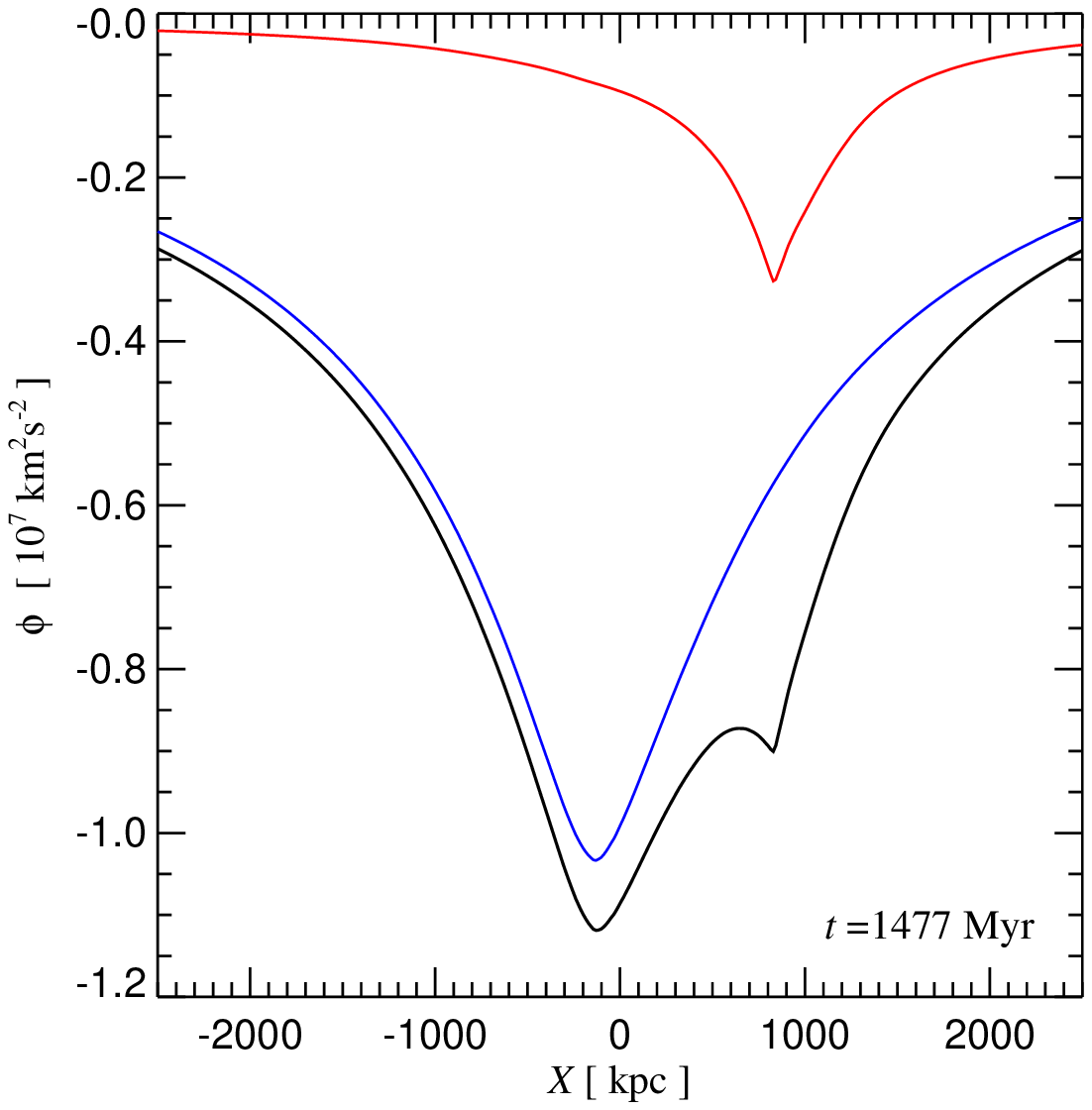}}\\%
\end{center}
\caption{The gravitational potential along the symmetry axis of the merger,
  shown at different times. In each panel, the red line (top-most line) shows
  the potential generated by the mass belonging initially to the infalling
  sub-cluster, while the blue line gives the potential generated by the parent
  clusters mass. The black line shows their sum (bottom-most line). The panel
  at $t=1340\,{\rm Myr}$ additionally contains the potentials generated by the
  original mass distributions of bullet and parent (dashed lines),
  horizontally translated to their current positions. The difference in the
  two total potentials arises from rearrangements in the mass distributions of
  the two components.}
\label{FigPotEvolv}
\end{figure*}

We note that the opening angle of the Mach cone seen in our simulations is
much wider than expected in the simple situation of a perturber moving with
constant supersonic velocity in a homogeneous medium that is at rest. This is
in part a result of the velocity structure of the upstream gas, which is not
at rest, a point we shall return to later on. In addition, the speed of the
bullet is not constant, and the temperature field is not strictly homogeneous.
As a combined result of these complications, the opening angle of the bow shock
cannot easily be used to obtain an independent measure of the strength of the
shock.

\subsection{The velocity of the bullet}

An important point to note about Figure~\ref{FigProfiles} is that the
shock velocity itself does not give the expected jump in the velocity
of the gas across the shock. The latter is only \be \Delta v= v_1- v_2
= v_s \left(1-\frac{\rho_1}{\rho_2}\right) = v_s \, \frac{2 ({\cal
M}^2-1)}{(\gamma+1){\cal M}^2} = 2937\,{\rm km\,s^{-1}},
\label{eqdeltav} \ee as a result of the continuity equation. Here
$\rho_1$ and $\rho_2$ give the gas densities in front of and behind
the shock, and $v_1$ and $v_2$ are the corresponding velocities. If
the latter are measured in the shock's rest frame, we have
$\rho_1\,v_1 = \rho_2\,v_2$, which leads together with the
Rankine-Hugoniot jump condition for $\rho_1/\rho_2$ to
Equation~(\ref{eqdeltav}).

But perhaps the most noteworthy feature in Figure~\ref{FigProfiles} is
the fact that the velocity of the gaseous bullet, which is located at around
$\sim 400\,{\rm kpc}$, is not at all equal to the shock velocity
$v_s$, as has often been assumed. Furthermore, our results show that the mass
centroid of the subcluster is in fact moving much slower; its velocity
lies around $\sim 2600\,{\rm km\,s^{-1}}$, yet the inferred shock
speed is $v_s = 4488\,{\rm km\,s^{-1}}$. The difference originates
primarily for two reasons:
\begin{enumerate}
\item The pre-shock gas is not at rest. Instead, it falls towards the
  bullet, reaching nearly $v_{\rm infall}=1100\,{\rm km\,s^{-1}}$ when
  it encounters the shock. This is because the gravity of the incoming
  bullet is felt already {\em ahead of the shock front}, unlike
  hydrodynamical forces.  We shall discuss this point further
  below. For now we note that this means that the shock front moves
  only with a velocity $\tilde{v_s} = v_s - v_{\rm infall}$ in the
  rest-frame of the system.
\item The subcluster's mass centroid and the X-ray centroid are not
  moving as fast as the shock front itself, and are, in fact,
  increasingly falling behind. Note that the post-shock gas has only a
  velocity of $\Delta v$ relative to the pre-shock gas, and not the
  full speed $v_s$ of the shock. The bullet will hence experience
  significant braking by ram-pressure if it moves faster than $\Delta
  v - v_{\rm infall}$ in the rest-frame, which it does.  This suggests
  that a difference between the speed of the bullet and the shock
  front may naturally arise, but the size of this effect is hard to
  estimate due to the time variability and the complicated geometry of
  the merger. But as we quantify below, a velocity difference between
  the subcluster's mass centroid and the bow shock is indeed present
  and amounts to $\sim 700\,{\rm km\,s^{-1}}$ at the time of best
  match.
\end{enumerate}

In Figure~\ref{FigSeparations}, we show the locations of the shock
front, the edge of the bullet, and the mass centroids of parent and
subcluster as a function of time, from a brief moment before core
passage over a period of 400 Myr. Clearly seen is that the shock front
moves faster than the mass centroid associated with the bullet. In
fact, numerical differentiation of these curves shows that the shock
front is moving with $\tilde{v}_s \simeq 3300\,{\rm km\,s^{-1}}$ in
the center-of-mass frame at time $t=1340\,{\rm Myr}$, while the mass
centroid is moving with $2600\,{\rm km\,s^{-1}}$. The edge of the
bullet stays closer to the shock front, however, only initially their
distance grows to $\sim 80\,{\rm kpc}$ but then stays roughly
constant.  Given the pre-shock infall of $1100\,{\rm km\,s^{-1}}$ of
the gas, the velocity of the shock front in the rest-frame of the
upstream gas is $\sim 4400\,{\rm km\,s^{-1}}$, consistent with the
shock velocity $v_s = {\cal M}c_s$ inferred from the Rankine-Hugoniot
jump conditions.

\begin{figure}[t]
\begin{center}
\resizebox{8.2cm}{!}{\includegraphics{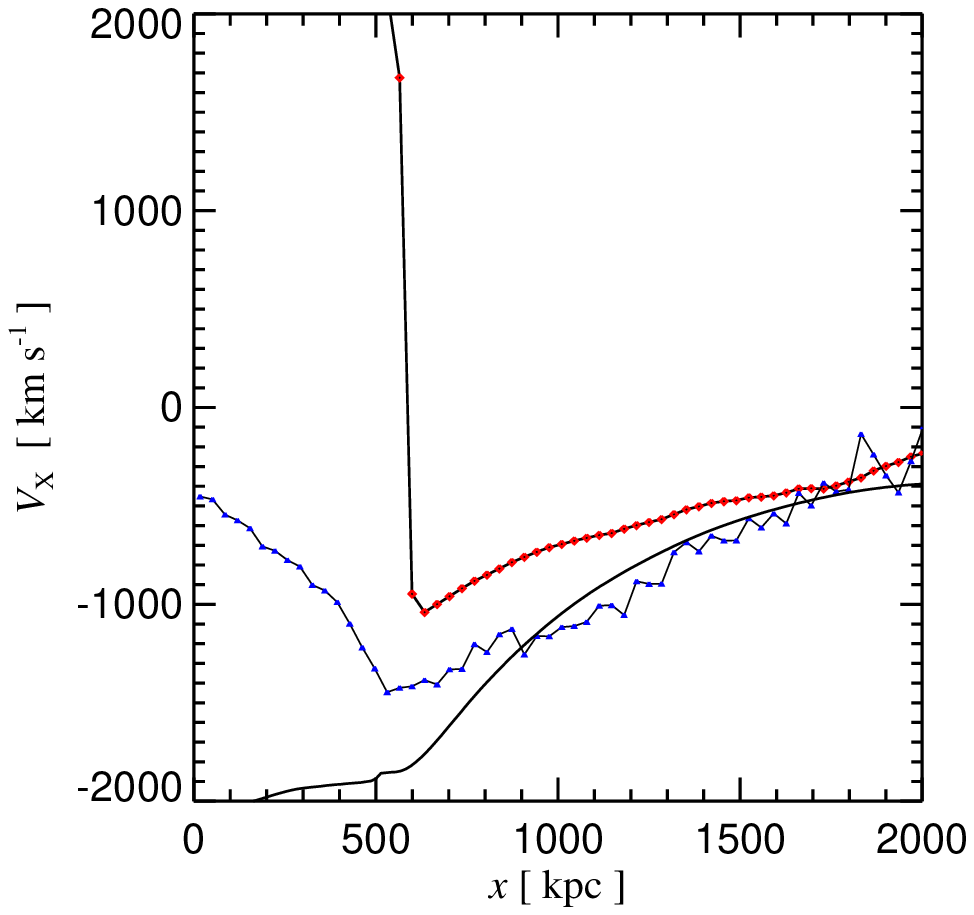}}
\end{center}
\caption{Gas infall velocity and mean streaming velocity of the parent
  cluster's dark matter along the symmetry axis of the system. The location of
  the bow shock is clearly seen as vertical discontinuity in the gas velocity.
  The solid black line shows an estimate for the mean dark matter velocity
  obtained from the change in the gravitational potential generated by the
  time variation of the mass distributions of parent cluster and infalling
  bullet.}
\label{FigInfallVel}
\end{figure}

We can also see that the core passage has happened $180\,{\rm Myr}$ before the
time displayed in Figure~\ref{FigProfiles}, at which point the mass peaks have
a separation of $627\,{\rm kpc}$. This is a bit smaller than the $720\,{\rm
  kpc}$ cited for the separation of the mass centroids of 1E0657--56, based on
gravitational lensing analysis. However, just $60\,{\rm Myr}$ later, the
distance grows to this level, with an unchanged shock strength, so that the
time of best match using this measure could be defined as $t\sim 1400\,{\rm
  Myr}$.

Comparing with the positions of the hydrodynamical features in
Figure~\ref{FigSeparations}, we see that for the chosen simulation
parameters the mass peak of the bullet lies slightly ahead of its
X-ray edge at the time of best match, at which point it is in fact
also quite close to the location of the shock front. While this is
broadly consistent with the results from weak- and strong lensing
analysis of 1E0657--56 \citep{clowe04,clowe06,bradac06}, the size of
the offset in this simulation is clearly smaller than inferred
observationally. We note that this is also the case for the offset
between the X-ray emission and mass peak associated with the parent
cluster: those features are quite well aligned in our simulation
model, unlike in the observations. However, we will consider a small
variation of our default simulation model later on in \S\ref{SecMatch}
where a more accurate match to the observational data is obtained.

\subsection{Gas infall towards the bullet}

In the previous section we have seen that the upstream gas ahead of
the bullet's shock front is not at rest relative to the main parent
cluster. Rather it is infalling towards the incoming bullet with a
significant velocity. This effect contributes substantially to the
large difference between the inferred shock velocity and the actual
velocity of the infalling subcluster. Here want to examine more
closely how this velocity arises.

In Figure~\ref{FigPotEvolv}, we show the time evolution of the gravitational
potential along the symmetry axes during the merger. Each panel shows the
potential at a different time, both for the total mass and separately for the
material that originally defines each of the two clusters before they start
overlapping. In the early phases of the collision, the mass distribution of
each of the clusters changes relatively little, such that the total potential
is approximately given by translating the initial potentials of the two
components, and summing them up. However, during core passage the summed
potential changes rapidly in time, leading to a phase of violent relaxation
that reshuffles the energies of individual dark matter particles. This
irreversibly modifies the mass distributions of the involved components, and
hence their potentials.  Recall that in a static gravitational potential the
specific energies of individual dark matter particles, $e_i = v_i^2/2 +
\Phi(\vec{x}_i)$, are constant along their orbits $\vec{x}_i(t)$, but if the
potential varies in time, their energies evolve at a rate ${\rm d}e_i/{\rm
  d}t = \partial\Phi/\partial t$, i.e.~the particle energies simply change
with the
rate of change of the potential at their current location. It is precisely
during core passage when the potential fluctuates on a very short timescale
such that a widespread rearrangement of particle energies can occur.

We now relate this phenomenon to the large infall velocity we have found in
the gas ahead of the shock. We argue that this occurs largely as a response to
the change in the gravitational potential of the system induced during core
passage. Consider to this end the panel at time 1340 Myr in
Figure~\ref{FigPotEvolv}. In addition to the direct measurement of the total
potential (solid lines), we have also included an estimate of the potential
obtained by linearly translating the potentials of the two original mass
components to the current positions of the mass centroids (dashed lines), and then
adding them. In the resulting comparison, it is seen that the dynamically evolved
potential has clearly become deeper during core passage relative to that
expected for unaltered mass distributions in the two components.  Note in
particular that the total potential in the infall region of the gas ahead of
the shock front has become deeper as well; this implies that the gas in this
region is accelerated towards the shock considerably more strongly than
predicted in a model where the dynamical response of the parent and bullet
during core passage is neglected.

We can use the above measurement of the `excess change' $\Delta\Phi$ in the
gravitational potential to obtain a rough estimate of the induced dark matter
streaming velocity in the parent's cluster material in the infall region ahead
of the shock. To this end we invoke energy conservation for the radial
streaming velocity of the dark matter, which at a given location should
roughly increase as $\tilde{v}_r^2 \simeq v_r^2 -2\Delta\Phi$. This estimate
neglects the fact that the particles really have a velocity distribution
function, but should be good for an order of magnitude estimate. In
Figure~\ref{FigInfallVel} we show the result, and compare it with the real
dark matter streaming velocity measured in the simulation, and the mean gas
velocity. We see that the above estimate matches the measurement
quite well. The gas inflow velocity is somewhat lower than the induced dark
matter streaming. This can be understood as a result of gas pressure
effects that resist the adiabatic compression associated with the induced
infall, and which are absent for the collisionless dark matter.

\section{Improving the match to 1E0657--56} \label{SecMatch}

\begin{figure*}[t]
\begin{center}
\resizebox{7.6cm}{!}{\includegraphics{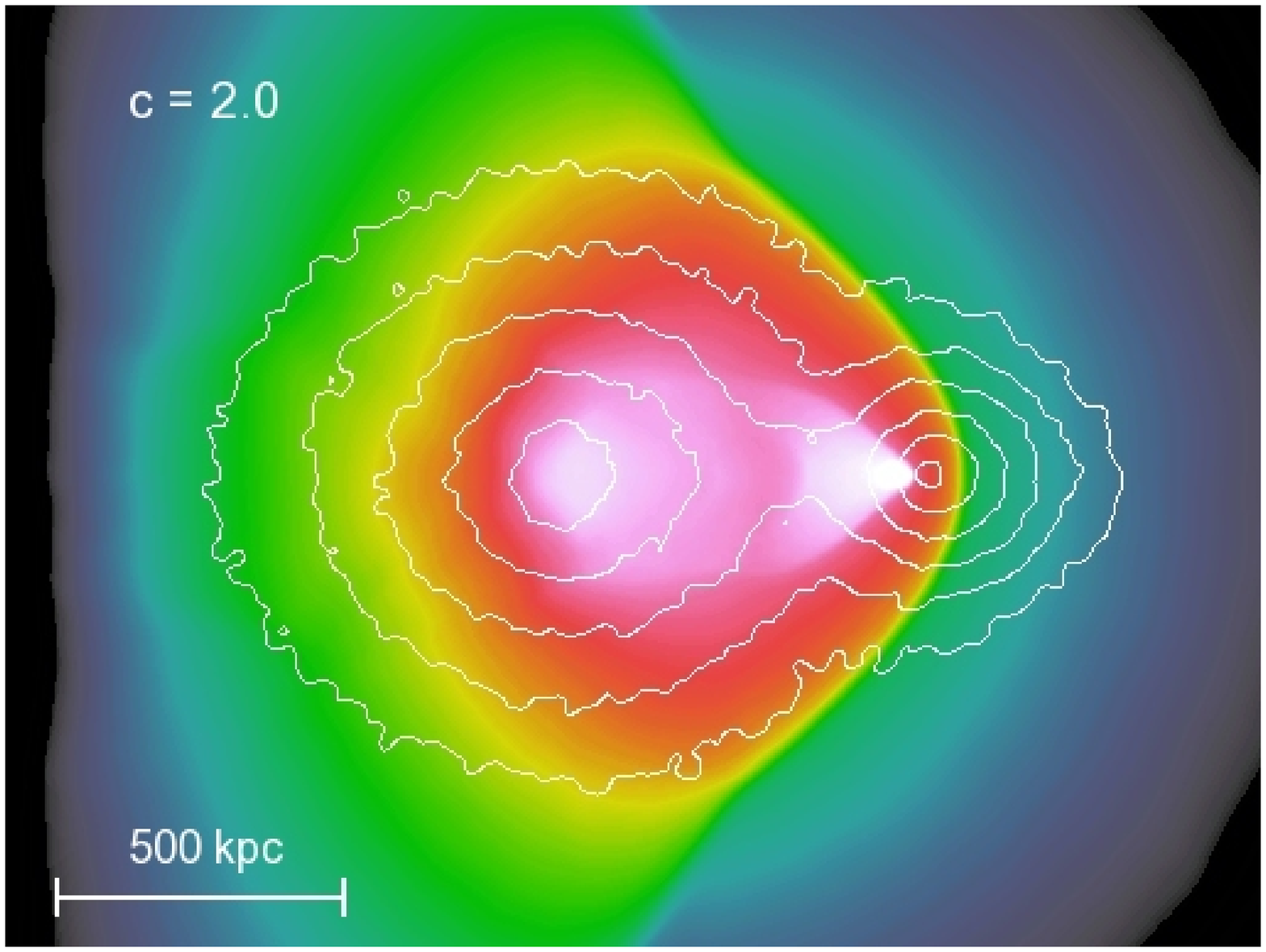}}\ %
\resizebox{7.6cm}{!}{\includegraphics{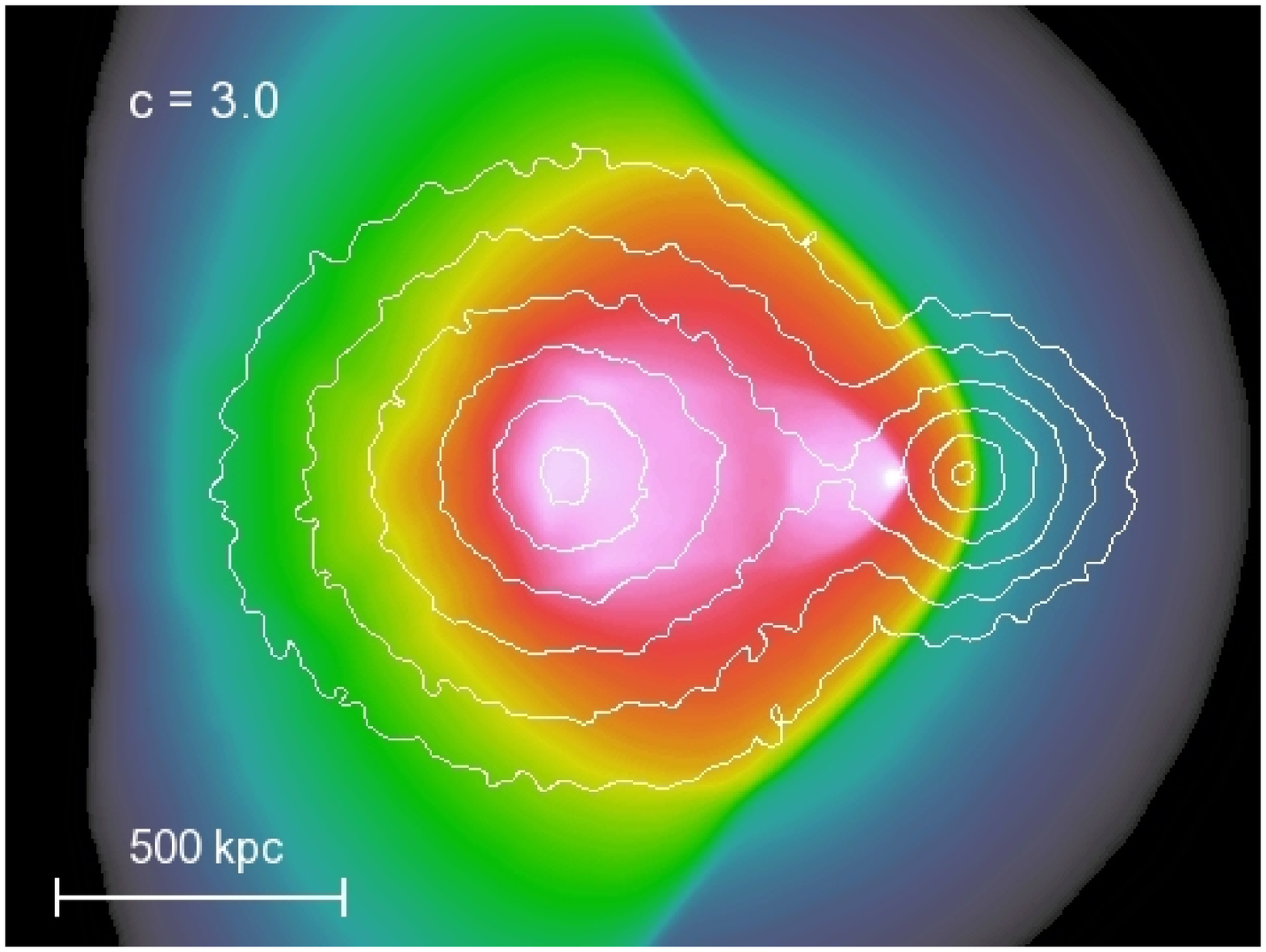}}\\
\resizebox{7.6cm}{!}{\includegraphics{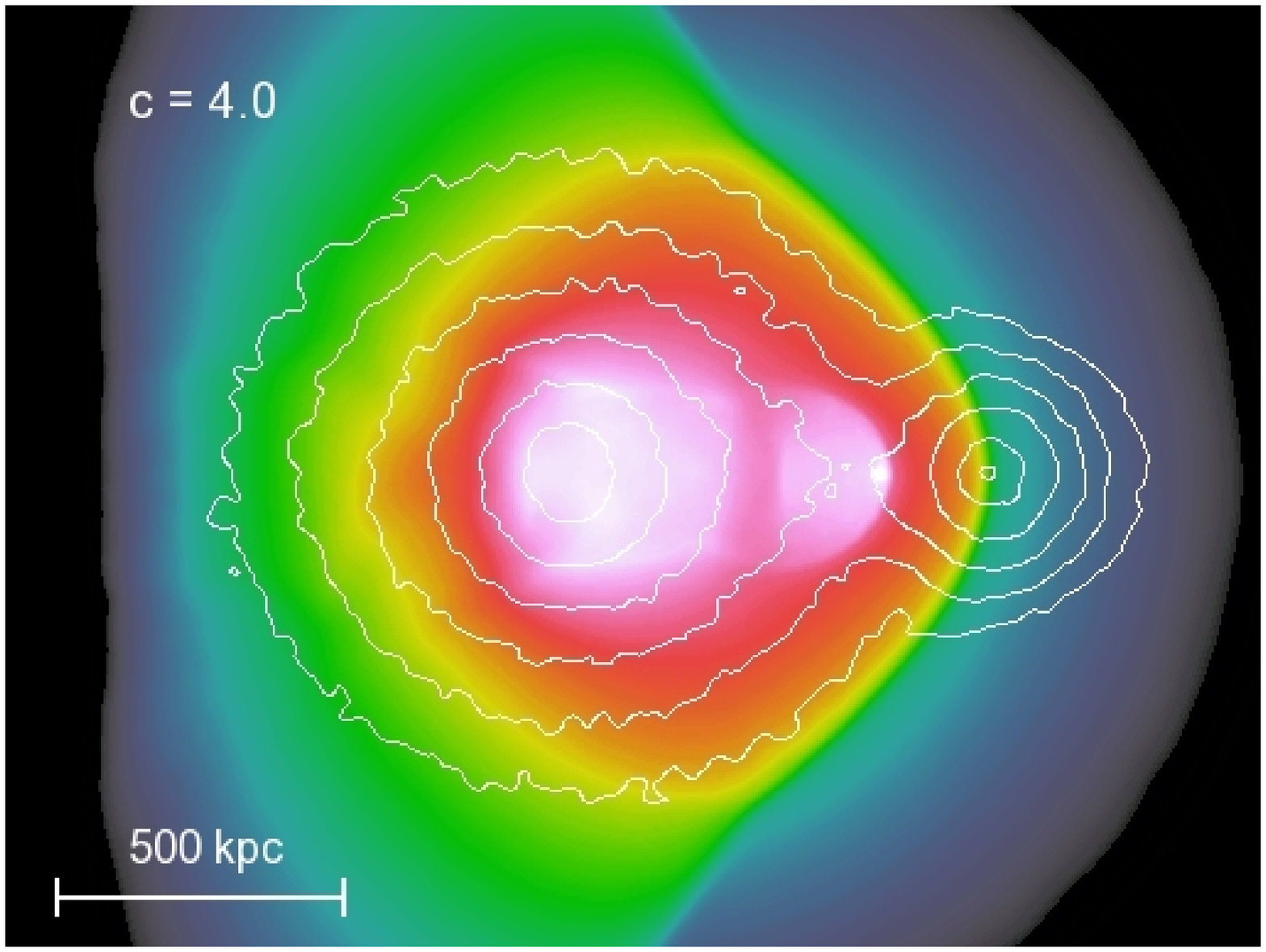}}\ %
\resizebox{7.6cm}{!}{\includegraphics{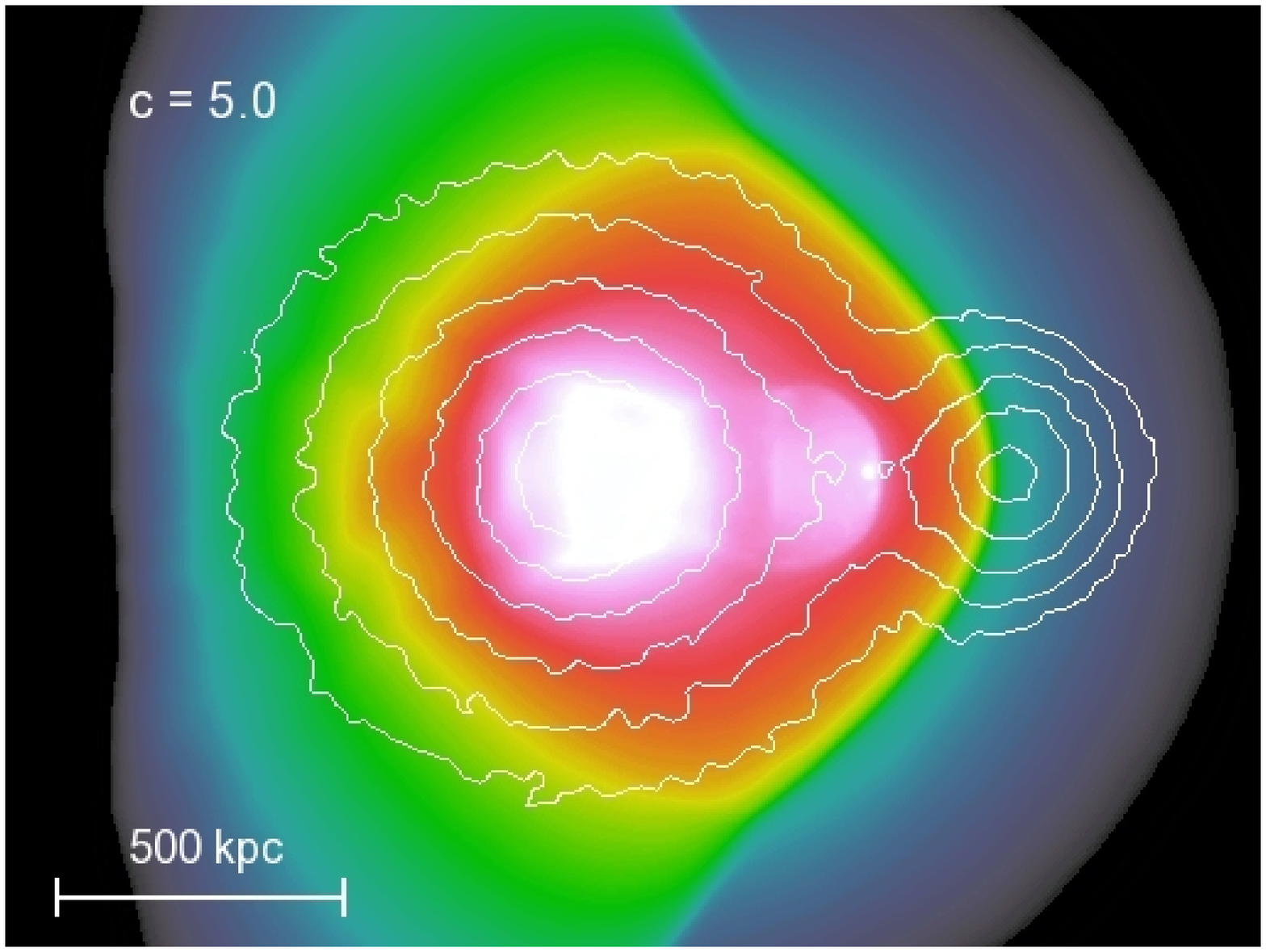}}\\
\end{center}
\caption{X-ray surface brightness and projected mass maps (contours)
  for simulations with different concentrations of the main cluster,
  as labeled.}
\label{FigXrayContours}
\end{figure*}

We now discuss in more detail the quality of the match of our default merger
model to the observations of 1E0657--56, and in particular consider the
sensitivity of our results to the assumed concentration of the parent cluster.

In Figure~\ref{FigXrayContours}, we compare maps of the X-ray emission
with overlaid contours of the projected total mass for different
choices for the initial concentration of the parent cluster prior to
the merger. The value of $c=2.0$ adopted in our default model (as
suggested by new mass models of Clowe matched to the lensing data) is
quite low for a cluster of this mass, given that the typical
concentration expected in $\Lambda$CDM models for a cluster of this
mass lies around $c\sim 4-5$ \citep{Eke2001,Bullock2001}.  We
therefore calculated a number of simulations with higher
concentrations, using $c=3$, $4$, and $5$.  A larger concentration of
the parent system will make it more difficult for the gas bullet to
penetrate through the center. In particular, we expect that the
incoming bullet experiences a higher ram pressure and will presumably
also be stripped more efficiently of its dark matter during core
passage.

The results shown in Figure~\ref{FigXrayContours} confirm this expectation:
With increasing concentration, the separation between the X-ray bright region
of the bullet (where the gas density peaks) and its associated dark matter
mass peak becomes larger.  In addition, the shape of the contact discontinuity
(the edge of the bullet) changes noticeably as well; it becomes progressively
rounder with increasing concentration, such that the opening angle of the
narrow wedge obtained in the $c=2.0$ case is first enlarged, and finally
becomes transformed into a round, blunt shape.  On the other hand,
the shape of the bow shock remains essentially unaffected by the change in
concentration.

\begin{figure*}[t]
\begin{center}
\resizebox{8.1cm}{!}{\includegraphics{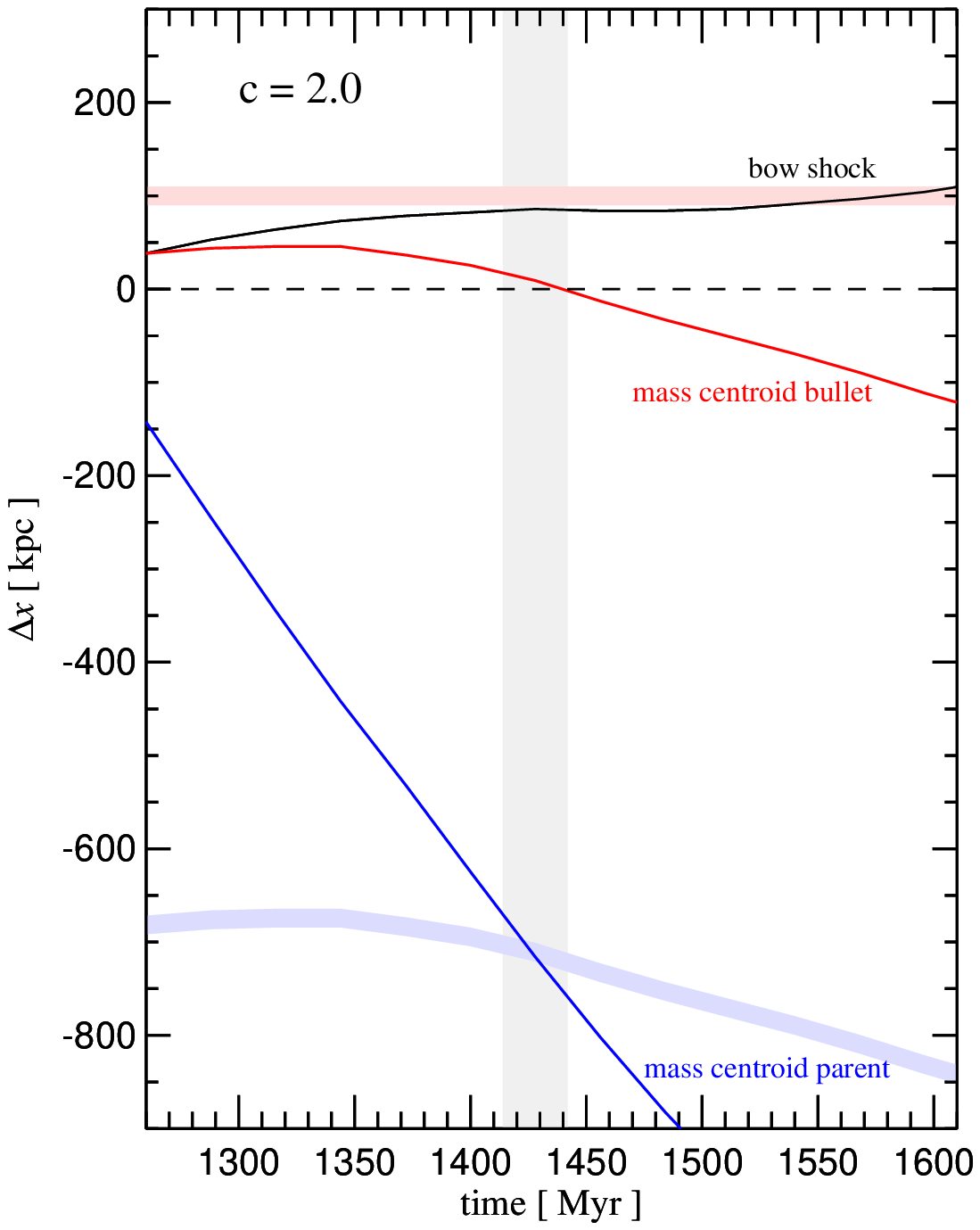}}\ %
\resizebox{8.1cm}{!}{\includegraphics{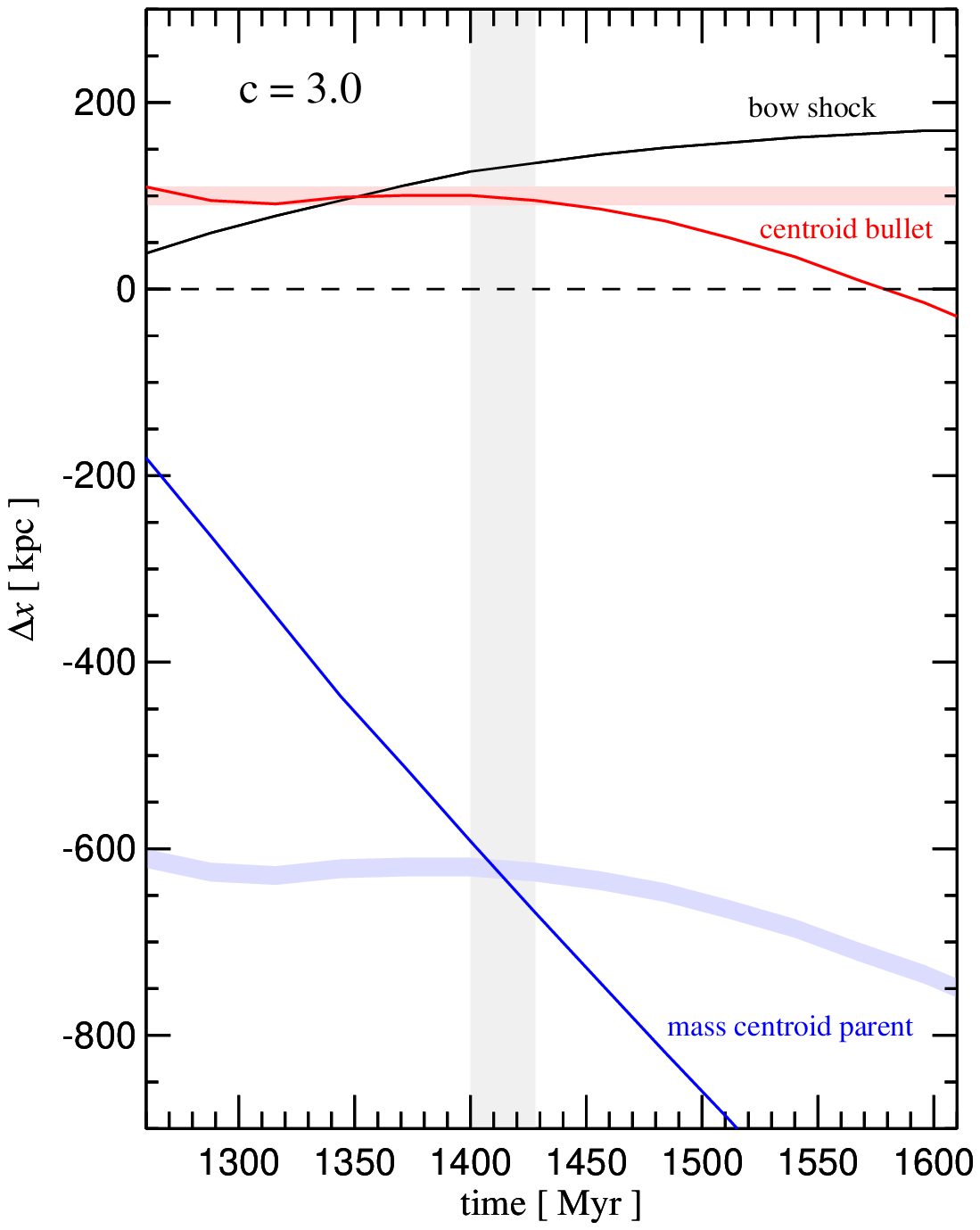}}\\
\end{center}
\caption{Relative distances between the mass-centroids of bullet and
  parent cluster, and the position of the bow shock, relative to the
  edge of the bullet (i.e. the cold front), as a function of time.
  Adopting a separation of 720 kpc for the distance of the mass
  centroids, and 100 kpc for the offset of the bullet's mass peak
  relative to the edge of the bullet, we can identify the time of the
  best match of the simulation to the observations. A simultaneous
  match of the two target separations is obtained for the $c=3$ model
  at time $t\simeq 1420\,{\rm Myr}$ (see shaded target regions). The
  default model with a lower concentration of the parent cluster of
  $c=2.0$ provides a much poorer fit, primarily because no significant
  offset between the position of the bullet's mass and its X-ray edge
  develops.}
\label{FigSeparationMatch}
\end{figure*}

By visual inspection, it appears that the $c=3.0$ case among this set
of simulations can best match the observed shape of the bullet's edge
in 1E0657--56. For this simulation we also find a significant spatial
offset between the mass peak and the X-ray emission, quite similar in
size to that found in the gravitational lensing analysis of
\citet{clowe06}. Figure~\ref{FigSeparationMatch} quantitatively
confirms this.  In it we show the location of the edge of the bullet,
the bow shock, and the two mass peaks, as a function of time, both for
the default case of $c=2.0$ and the $c=3.0$ model. In order to
identify the time of best-match to the observed system, we plot the
positions relative to the coordinate of the edge of the bullet and we
mark the desired spatial offsets of the observed system with shaded
bands.  We focus on reproducing the $720\,{\rm kpc}$ distance on the
sky between the mass peaks of parent cluster and subcluster given by
\cite{clowe06}, and on the separation of $\sim 100\,{\rm kpc}$ between
the bow shock and the edge of bullet.  For a good match, we hence
require that these two relative differences are reproduced
simultaneously. We see that this is quite well the case for the $c=3$
simulation, at around $t=1420\,{\rm Myr}$.  Our default $c=2$
simulation underpredicts the separation between the bullet's mass peak
and edge of the bullet at all times, while the $c=4$ and $c=5$ cases
(not shown) overpredict it during the times when the mass peaks have
the right separation.

\begin{figure}[t]
\begin{center}
\resizebox{8.5cm}{!}{\includegraphics{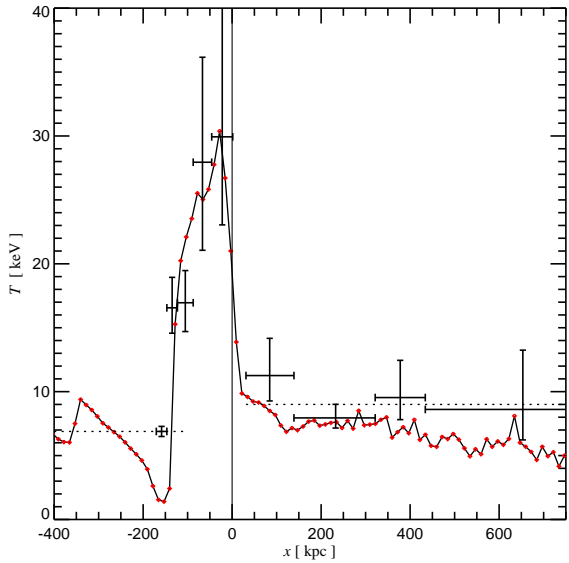}}
\end{center}
\caption{Comparison of the temperature profile across the shock front with
  deep Chandra observations of 1E0657--56. The shock-front is marked with a
  vertical line. The crosses are the measurements by \citet{markevitch06},
  with the leftmost point giving the temperature of the bullet.}
\label{FigTempProfile}
\end{figure}

It is interesting that the $c=3$ case also reproduces other existing
observational constraints on 1E0657--56 quite well.  In
Figure~\ref{FigTempProfile}, we show an enlargement of the
simulation's temperature structure across the shock front and the
contact discontinuity, and compare it to measurements of the
temperature profile of 1E0657--56 obtained by \citet{markevitch06}
based on deep Chandra data. The shock seen in the $c=3.0$ simulation
has a Mach number of ${\cal M}=2.9$, consistent with the observational
data. The fall off of the temperature at the cold front is reproduced
quite well. Also, the pre-shock temperature of 9 keV is matched well
by the model; only further ahead of the shock the simulated
temperature appears marginally lower than the measurements.

We can also consider the gas masses in spherical apertures of $100\,{\rm kpc}$
around the mass centroids of bullet and parent cluster, and about the center
of X-ray emission of the bullet. These masses have been measured by
\citet{clowe06}, who estimate $(2.7\pm 0.3)\times 10^{12}\,M_\odot$ of gas
around the bullet's mass centroid (actually they used the BCG position, which we take
here as a proxy for the mass centroid), while the simulation gives $1.5\times
10^{12}\,M_\odot$.  For the parent mass centroid they get $(5.5\pm 0.6)\times
10^{12}\,M_\odot$, and the simulation yields $6.1\times 10^{12}\,M_\odot$.
Finally, for the bullet's X-ray centroid they find $(5.8\pm 0.6)\times
10^{12}\,M_\odot$ while the simulation model gives also $5.8\times
10^{12}\,M_\odot$. We note however that unlike found in the lensing mass
reconstructions of \citet{clowe06} and \citet{bradac06}, our simulation does
not show a significant spatial offset between the mass centroid of the parent
cluster and its associated X-ray centroid.

\begin{figure}[t]
\begin{center}
\resizebox{8.5cm}{!}{\includegraphics{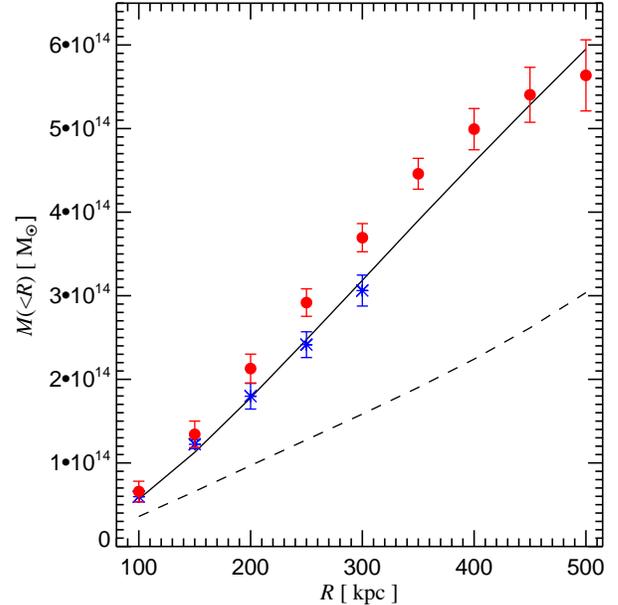}}
\end{center}
\caption{The cumulative mass profile of our simulation model compared to
  gravitational lensing results for 1E0657--56. The filled circles give the
  mass profile obtained by \cite{bradac06} for the parent cluster based on a combined
  weak and strong lensing analysis. The stars mark their inferred masses for
  the subcluster component.}
\label{FigCumulMassProfile}
\end{figure}

In Figure~\ref{FigCumulMassProfile}, we consider the projected cumulative mass
profile as a function of radius around the bullet's and the parent's mass
centroids, and we compare this to the measurement of \citet{bradac06} based
on their weak- and strong lensing reconstruction. Our model matches the result
of the lensing analysis for the parent cluster very well, but predicts
significantly less enclosed mass for the bullet itself. This could perhaps be
remedied by simulations with a more massive, a more concentrated, or a
non-spherical bullet.  Alternatively, it could be due to a systematic
overestimate in the lensing mass estimate of the bullet. While the latter
seems unlikely due to the extensive tests that have been applied to the mass
reconstruction methods applied to 1E0657--56, the fact that the baryonic masses
of the simulation model agree well with the directly measured gas masses from
the X-ray observations is a point supporting the simulation's mass model.

Finally, an interesting observation about the shape of the bow shock in our
simulated mergers is that it features a clear bend towards a larger opening
angle, approximately $600\,{\rm kpc}$ away from the symmetry axis of the
merger.  This feature originates during the core passage of the bullet, when
the central part of the bullet is accelerated to its peak velocity and shoots
out on the other side of the core.  It is presently unclear whether such a
feature exists in the observations, as it should be located in a region where
the available X-ray maps have extremely low surface brightness and are
dominated by noise. With some good will, one can spot a tentative hint for it
in the observed maps of \citet{markevitch06}, but this is certainly far from
being significant.  Future X-ray observations should be able to probe to
fainter surface brightness levels, in which case this generic prediction of
our simulation models could be tested.

\subsection{The future evolution of the bullet cluster}

It is interesting to use our simulation model to study the expected
future evolution of the bullet cluster system. Adopting the $c=3$
simulation as our best-matching model, we predict that the bullet will
continue to enlarge its distance from the parent cluster for another
$900\,{\rm Myr}$ from the time of best-match. After turn around, it
falls back and experiences a second core passage $2500\,{\rm Myr}$
after the first one. This is followed by a third passage another
$1700\,{\rm Myr}$ later, after which the bullet is nearly completely
destroyed, such that a `ring down' with a quick succession of possibly
several further core passages leads to a complete coalescence of the
cores. This allows the cluster to reestablish hydrostatic equilibrium,
so that we can expect 1E0657--56 to reach a fully relaxed state again
within the next $\sim 4\,{\rm Gyr}$.

\begin{figure}[t]
\begin{center}
\resizebox{8.5cm}{!}{\includegraphics{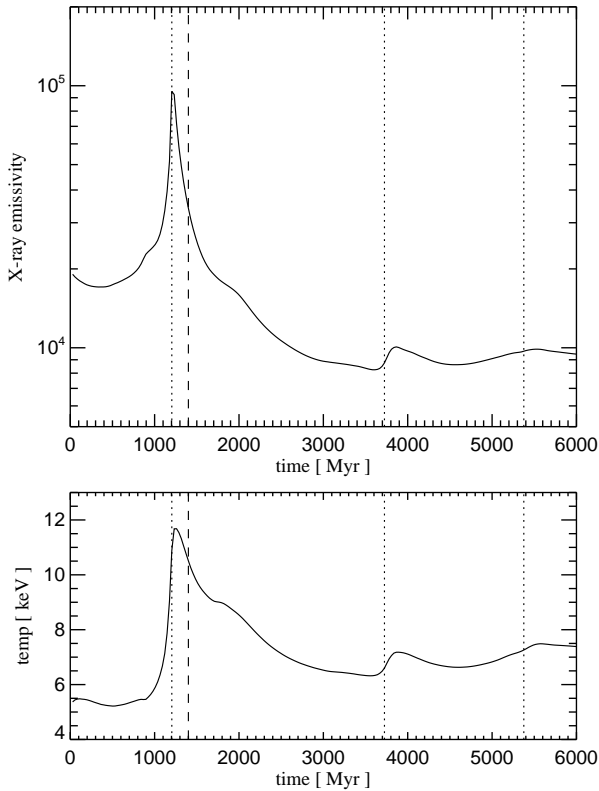}}
\end{center}
\caption{The time evolution of total X-ray emissivity (top panel) and
  mean gas temperature (bottom panel) of our best-matching simulation
  model for 1E0657--56.  The time corresponding to the observed moment
  is marked with a vertical dashed line, while the dotted lines give
  the core passages of the bullet.}
\label{FigXrayTimeEvolution}
\end{figure}

In Figure~\ref{FigXrayTimeEvolution}, we show the time evolution of
the expected bolometric X-ray emission of the system until the merger
is complete.  For simplicity, we have simply computed the quantity
$L_{\rm X} \propto \int \rho_{\rm g}^2 \sqrt{T} {\rm d}x$ as a proxy
for the bremsstrahlung emissivity from the gas of the simulations,
without taking into account metal lines. The vertical dotted lines
mark core passages of the bullet cluster. These happen slightly
earlier than the induced peaks in the X-ray emissivity. At the time of
best match with the observed state of 1E0657--56 (marked with a dashed
line), the system is in an {\em overheated state}, with a mean
emission-weighted temperature of around $11~{\rm keV}$. The cluster
responds to the compression from the merger by adiabatically
overexpanding past its equilibrium in the subsequent evolution,
leading to a comparatively cool state for a period of order $1\,{\rm
Gyr}$ before the second collision. Then a further heating event occurs
when the surviving part of the bullet passes through the core a second
time. Finally, the third core passage leads to near complete
disruption of the remains of the bullet.  We are then left with a
relaxed cluster with a mean emission-weighted temperature of $\sim
7.5\,{\rm keV}$.

\section{Dependence on model parameters} \label{SecParamDep}

\subsection{Orbital angular momentum of the encounter}

\begin{figure}[t]
\begin{center}
\resizebox{7.5cm}{!}{\includegraphics{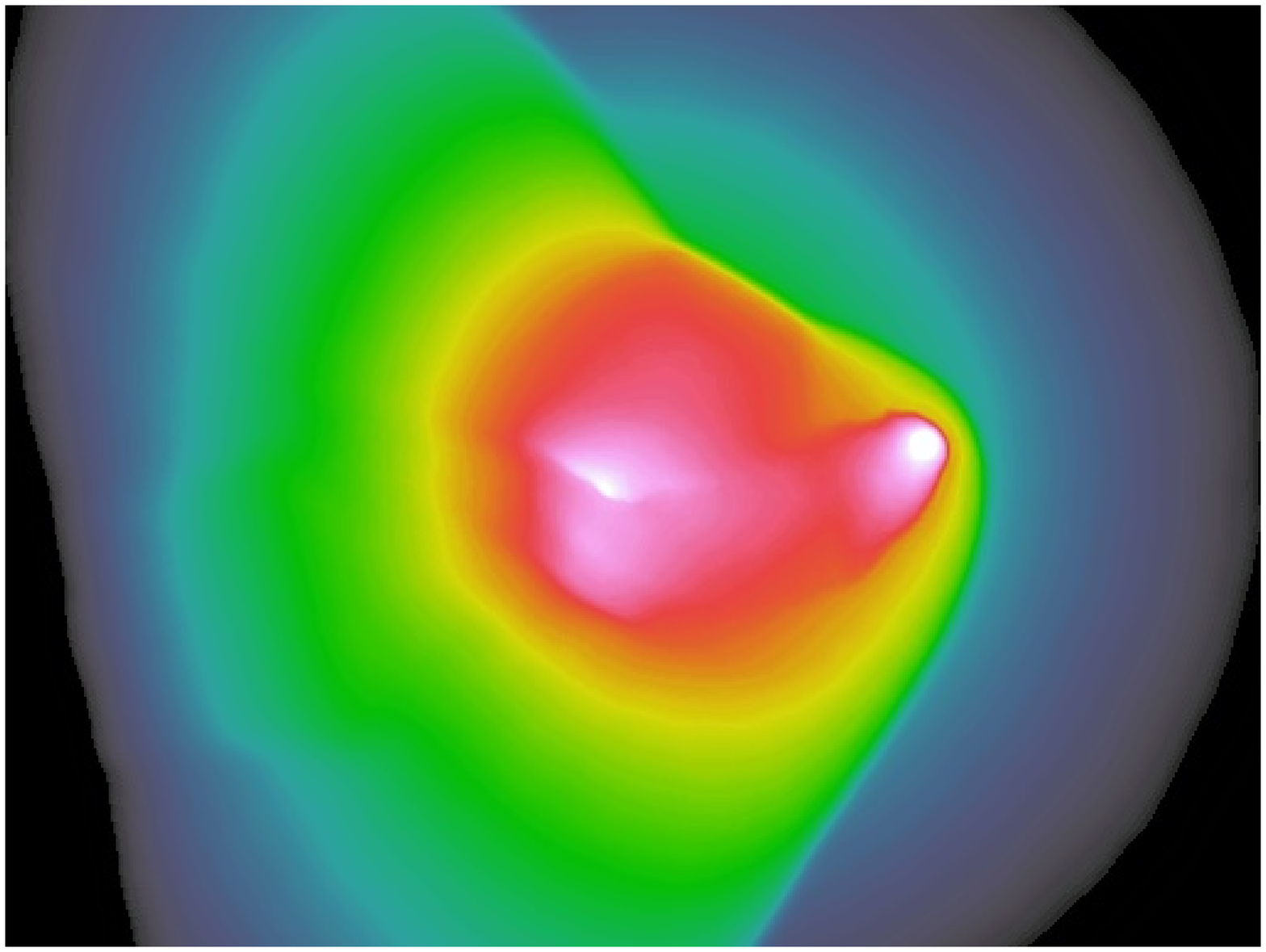}}\\
\resizebox{7.5cm}{!}{\includegraphics{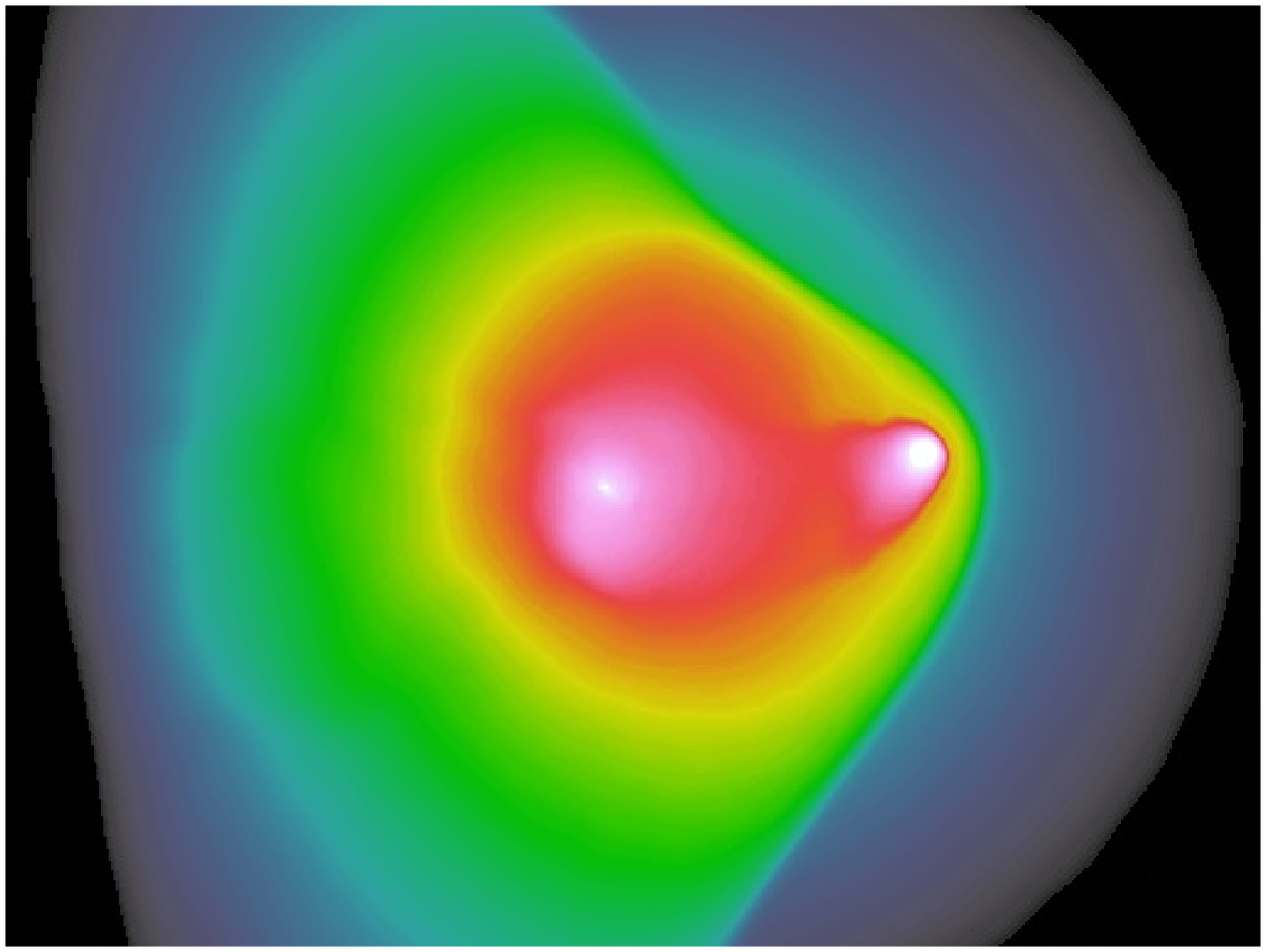}}\\
\resizebox{7.5cm}{!}{\includegraphics{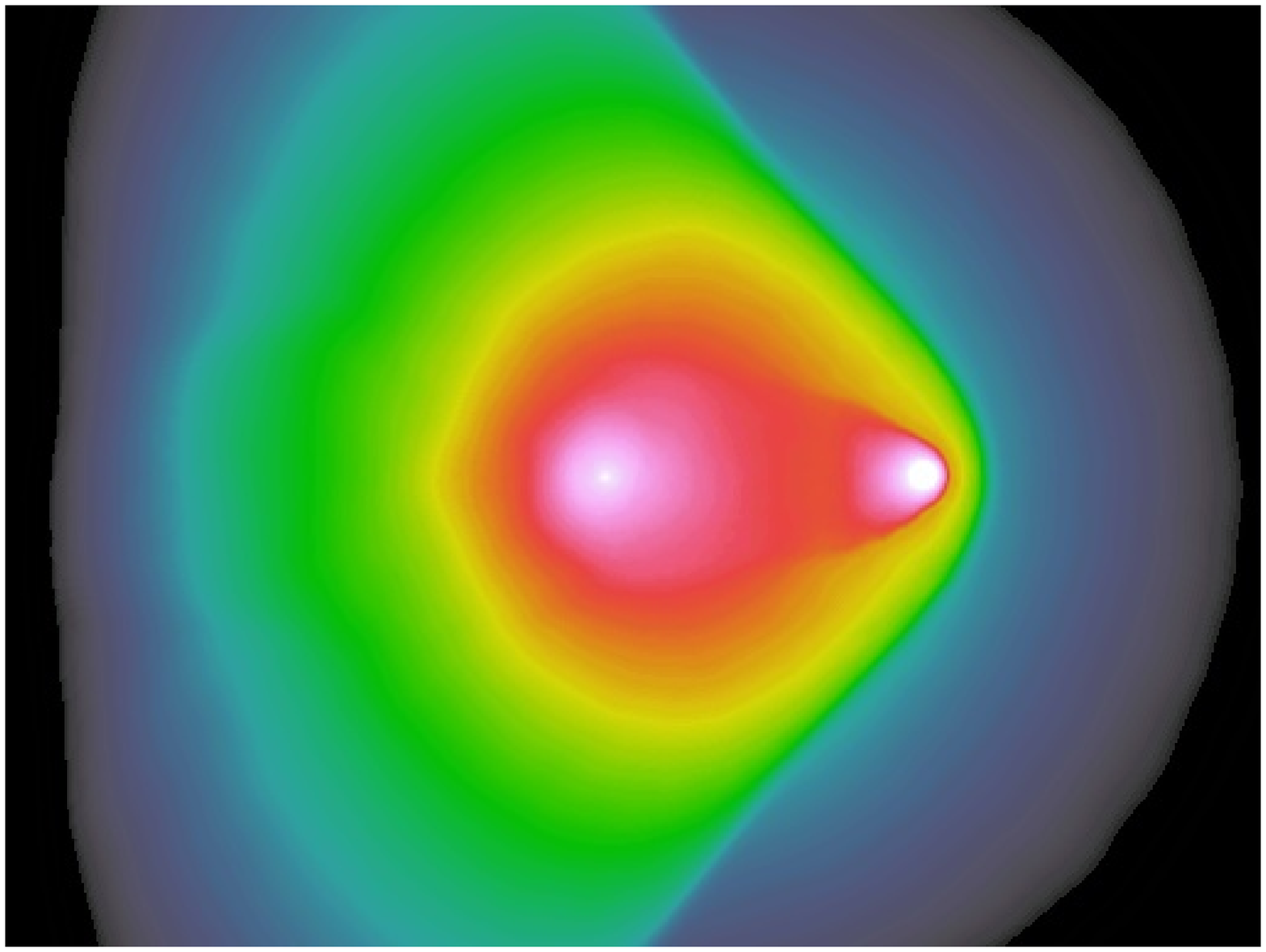}}\\
\end{center}
\caption{X-ray surface brightness profiles for a variation of our default
  merger model where the encounter occurs with impact parameter $b=12.5\,{\rm
    kpc}$. The panel on top shows the system with a face-on orientation of the
  orbital plane. In the middle panel, the orbital plane is at 45 degrees with
  respect to the projection plane, in the bottom panel it is seen edge-on.}
\label{FigNonCentralMerger}
\end{figure}

Based on the visual appearance of 1E0657--56 and its small line-of-sight
velocity difference of $\sim 600\,{\rm km\,s^{-1}}$ \citep{barrena02}, it
appears that the merger is very nearly central, and closely aligned with the
plane of the sky. Indeed, a slight offset from a central collision can easily
introduce a significant degree of axial asymmetry larger than observed in the
cones of bow shock and cold front. We illustrate this in
Figure~\ref{FigNonCentralMerger}, where we show in the top panel a projected
X-ray map for a merger with small orbital angular momentum, parameterized by
an `impact parameter' $b=12.5\,{\rm kpc}$.  Here $b$ is the distance at
pericenter the clusters would reach if they interacted as point masses during
the collision. Comparing this result with the observed X-ray map in
Figure~\ref{FigXrayMatch}, we see that the shock structure is already too
asymmetric to provide a good match to the observed X-ray morphology.

However, we can always recover symmetry by turning the orbital plane from
observing it face-on to observing it edge-on. In the middle and bottom panels
of Figure~\ref{FigNonCentralMerger} we show the same simulation when viewed
45~degrees and 90~degrees with respect to the orbital plane. While in these
cases symmetry is partially/fully recovered, the edge of the bullet and the
bow shock appear rounder and less sharp, and the edge of the bullet develops a
shape that agrees comparatively poorly with the observed one.  The
line-of-sight velocity differences obtained in the two cases amount to
$235\,{\rm km\, s^{-1}}$ (for 45 degrees) and $340\,{\rm km\, s^{-1}}$ (for 90
degrees), respectively, which means that one will quickly run into tension with
the observations when one tries to `hide' a larger orbital angular momentum in
the encounter than considered here with an edge-on view of the system.

We hence conclude that 1E0657--56 should be quite close to a central
merger that happens in the plane of the sky, even though a small
deviation from this geometry appears likely. For the sake of
simplicity, we treat all other merger simulations in this paper as
exactly central, with the merger axis aligned with the plane of the
sky.

\subsection{Baryon fractions of the clusters}

In our default models, we consider equal baryon fractions for the two merging
clusters, given by $f_b=0.17$, as suggested by recent cosmological parameter
constraints based on WMAP3 \citep{spergel06}, and we neglect the small
fraction of baryons locked up in stars, so that all of the baryons are
contained in the diffuse gas of the ICM. Clusters are generally believed to be
essentially closed-box systems due to their large mass, so the assumption of a
baryon fraction equal to the universal value should be comparatively safe.  On
the other hand, there is some tension with respect to this in the low-density,
low-$\sigma_8$ cosmological model favored by WMAP3, since detailed accounting
of all observed baryonic components in clusters tends to lead to values for
$f_b$ smaller than 0.17 \citep{McCarthy2006}.

However, in a number of test simulations where we varied $f_b=0.17$,
we found that our results are relatively insensitive to changing this
parameter, unless we assign different baryon fraction to the
clusters. The latter might be appropriate if the smaller sub-cluster
has lost some of its baryons as a result of strong outflows in
progenitor systems, or has locked up more of its baryons into
collisionless stars.  In this case we expect trends that are closely
related to those found for changes in the concentration, which we
considered earlier. If the baryon fraction of the bullet is lowered
relative to that of the main cluster, it will be harder for the
bullet's gas to pass through the main cluster's core. This should
enlarge the separation between mass centroid of the bullet and the
X-ray peak after core passage. Conversely, if the bullet is made more
gas-rich while the parent's gas density is kept fixed or lowered, the
deceleration of the bullet by ram-pressure will become smaller,
similar to the effect obtained when the concentration of the parent
cluster is lowered, which will also tend to keep the mass centroid of
the bullet and its X-ray cold front closer together.

We have carried out a number of additional simulations where we varied the
baryon fractions in the two merging systems, but otherwise adopted the
parameters of our default model. The results of these simulations
qualitatively confirm the expectations described above without improving the
match to the 1E0657--56 system in any significant way. For the sake of
brevity, we therefore omit showing X-ray maps for these runs in a separate
figure.

\subsection{Numerical resolution}

For all our default simulations we used a mass resolution of $m_{\rm
  dm}=6.225\times 10^{8}\, M_\sun$ and $m_{\rm gas} = 1.275\times 10^{8}\,
M_\sun$, translating to $4\times 10^6$ particles for the parent cluster and
$4\times 10^5$ for the infalling bullet cluster. To test for any possible
dependence on numerical resolution, we have rerun our default model with an
increased particle number by a factor of 8, and an improved spatial resolution
by a factor of 2 per dimension. This simulation produces results for the
strength of the shock and for the locations and velocities of the various
hydrodynamic and dark matter features that are in excellent agreement with the
simulation computed at our standard resolution.

We conclude form this resolution test that the numerical resolution used in
all our simulations is sufficient to provide numerically converged
results for all the quantities studied here.

\subsection{Other parameters}

Other physical parameters that in principle should be addressed in the
simulations if one wants to find a `perfect' model for the 1E0657--56 system
include the mass ratio of bullet and parent cluster, their intrinsic shape,
the relative central entropy levels of the gas in the clusters, their relative
baryon fractions, the effective viscosity of the gas \citep{Sijacki2006}, its
thermal conductivity \citep{Jubelgas2004}, and also non-thermal pressure
components like cosmic rays \citep{Prommer2006,Ensslin2006,Jubelgas2006} or
intracluster magnetic fields \citep{Dolag2002}. It would also be interesting
to try to identify a 1E0657--56 look-alike in a large-scale cosmological
hydrodynamical simulation, where the full dynamics of the infall and of gas
accretion would be treated self-consistently.

Obviously, this long list of possibilities quickly leads to a very large
parameter space which will be difficult to constrain without additional
observational constraints. Given that already our simple toy model provides a
good match to the key features that are observed in 1E0657--56, we refrain from
studying the  extended physical parameter space further in this paper,
apart from one exception that we shall consider in the next section.

\section{Mergers in fifth force models} \label{fifthforce}

Many extensions of the standard model of particle physics not only
predict dark matter candidates in the form of new elementary
particles, but also additional (long-range) fields which may couple
differently to the dark matter particles and ordinary baryons
\citep[e.g.][]{damour1990,farrar04,gubser2004}.  Interest in such
models is strong also on astrophysical grounds. For example, they can
produce cosmic voids that are `emptier' of galaxies than standard cold
dark matter models \citep[e.g.][]{farrar04,nusser2005}, which might be
required to match observational data.

One of the simplest classes of such models can be phenomenologically
characterized by a `fifth forth' in the dark sector, mediated by a
Yukawa-like potential \be \phi_s(r) = -\beta
\frac{G\,m}{r}\exp\left(-\frac{r}{r_s}\right) \ee between dark matter
particles of mass $m$, in addition to ordinary gravity. Here $r_s$ is
an effective screening length of the scalar force, while $\beta$ gives
its strength on small scales relative to ordinary gravity.  The
difficulty to explain a velocity of the bullet as high as $\sim
4700\,{\rm km\,s^{-1}}$ in $\Lambda$CDM cosmologies led
\citet{farrar06} to suggest that such a high speed, if true, may
provide direct evidence for the existence of a fifth force.

In this study, we have shown that the common approach of equating the shock
velocity with the bullet velocity is incorrect, and that $\Lambda$CDM models
in fact quite naturally produce shock speeds in excess of $4500\,{\rm
  km\,s^{-1}}$, despite the fact that the bullet is moving much slower than
that, by about $\sim 2000\,{\rm km\,s^{-1}}$. While this essentially
eliminates the utility of 1E0657--56 to obtain strong constraints on the
strength of a possible fifth force, we are here interested in explicitly
testing the conjecture of \citet{farrar06}, i.e.~we want to see by how
much a fifth force can accelerate the bullet in direct simulations of the
merging process.

For definiteness, we choose a range of $r_s=5.5\,{\rm Mpc}$ and explore two
different values of $\beta$, namely $\beta=1.0$ and $0.3$. The size of $r_s$
is large enough to encompass both the parent cluster's virial radius and that
of the infalling bullet cluster, such that the gravitational constant between
dark matter particles is effectively boosted to $(1+\beta)G$ when they are
isolated. If we hence increase the random dark matter velocities in the dark
halos of the initial clusters by a factor $[(\beta + 1+ f_b)/(1+f_b)]^{1/2}$,
we retain dynamical stability of the isolated systems in the fifth force
model, with the same density profile as in our default simulations.

The range $r_s$ is also short enough such that a parabolic orbit from
infinity leads only to a slightly increased velocity compared to a Newtonian
infall when the virial radii first touch. We approximate this initial velocity
of the encounter with the relative velocity two point masses would reach when
falling together from infinity up to the initial relative separation of the
sum of the two virial radii, which is the starting separation adopted in our
simulations. In doing so we implicitly assume that the relative binding
forces between gas and dark matter hold the isolated systems together during
the initial infall without appreciable shape distortions due to the violation
of the equivalence principle in the fifth force model. We then expect a 17.2\%
increase in the relative velocity at the initial time for the $\beta=1$ model,
which we correspondingly include in our initial conditions for the fifth force
simulation. We note that for an infinite range of the fifth force, this
correction factor would increase to 30\%.

\begin{figure}[t]
\begin{center}
\resizebox{8.5cm}{!}{\includegraphics{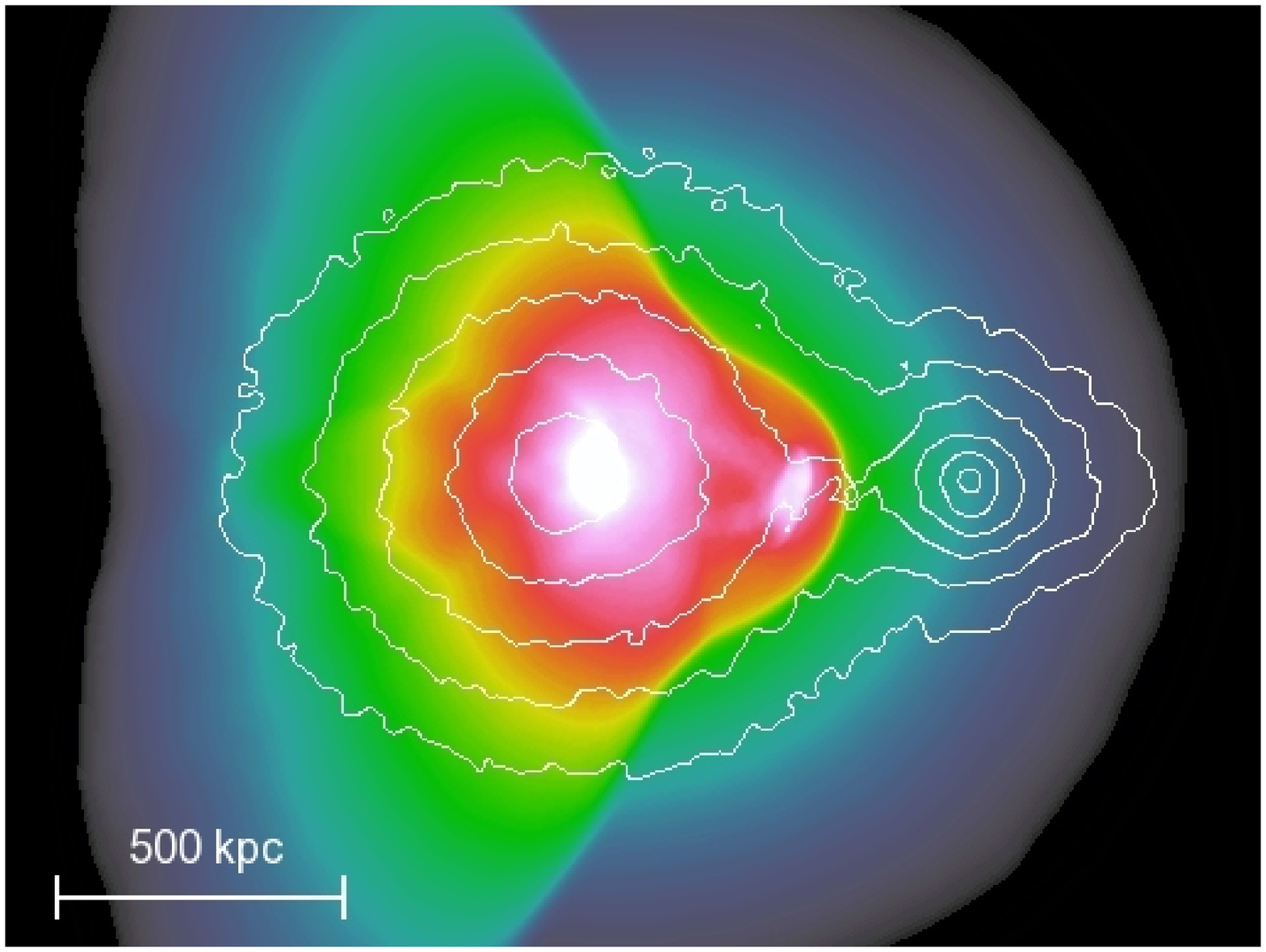}}\\
\resizebox{8.5cm}{!}{\includegraphics{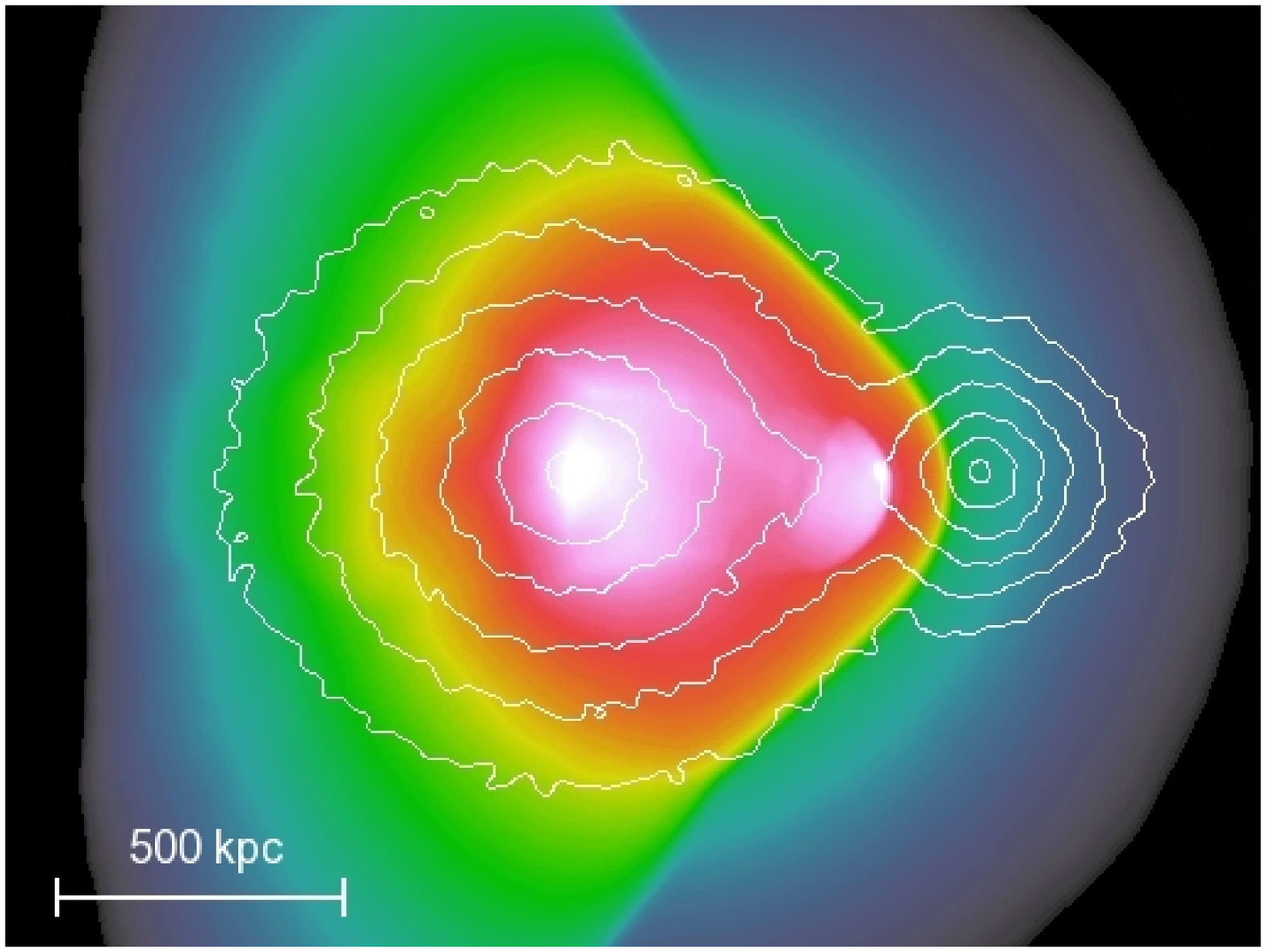}}
\end{center}
\caption{Map of the X-ray emission of a merger model with a 5th-force
  in the dark matter sector, characterized by $\beta=1.0$ (top panel)
  and $\beta=0.3$ (bottom panel).  The projected mass distribution is
  given by overlaid contours. In both cases, the simulations are shown
  at a time when the two mass peaks are separated by $720\,{\rm
  kpc}$. }
\label{FigXrayMassContWithFifthForce}
\end{figure}

In Figure~\ref{FigXrayMassContWithFifthForce}, we show maps of the
X-ray surface brightness with overlaid mass contours for the
simulations with $\beta=1.0$ and $\beta=0.3$, respectively, at a time
when the mass centroids of the bullet and the parent cluster are
$720\,{\rm kpc}$ apart. The X-ray morphology seen in the maps can be
directly compared to the corresponding maps for ordinary mergers shown
in Fig.~\ref{FigXrayContours}. What is immediately apparent is the
rather large offset between the dark matter and the gas of the bullet
in the $\beta=1.0$ case.  The centroid of the dark mass of the bullet
is $230\,{\rm kpc}$ ahead of the shock front in this case, and $300
\,{\rm kpc}$ ahead of the edge of the bullet. An analysis of the bow
shock shows that the Mach number of the shock is slightly lower in this
fifth force run, with ${\cal M} = 2.5$, but the preshock temperature
is a bit higher (10 keV), such that the inferred shock velocity is
$\sim 3900\,{\rm km\,s^{-1}}$.  Also in this case there is a preshock
inflow velocity, but with a slightly reduced size of $\sim 700\,{\rm
km\,s^{-1}}$, which in part is explained by the earlier time of the
`best-match', here defined in terms of the separation of the mass
centroids of bullet and parent cluster.

\begin{figure}[t]
\begin{center}
\resizebox{8.5cm}{!}{\includegraphics{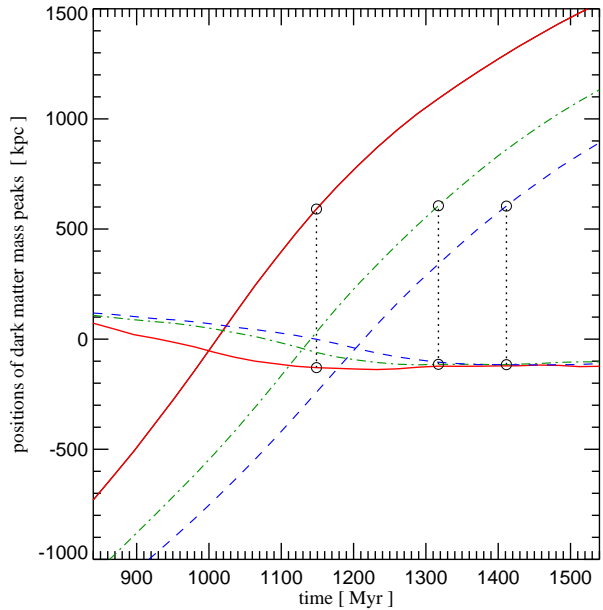}}
\end{center}
\caption{Positions of the dark matter mass peaks of bullet and parent
  cluster as a function of time. We compare a 5th-force model with
  $\beta =1.0$ (solid lines) and $\beta =0.3$ (dot-dashed lines) with
  our default model (dashed lines). The times where the separation of
  bullet and parent cluster are equal to $720\,{\rm kpc}$ are marked
  with dotted vertical lines. For the fifth force models, the bullet
  reaches a higher velocity, and core passage happens earlier. When
  the separation is $720\,{\rm kpc}$, the bullet's dark matter moves
  with $3800\,{\rm km\,s^{-1}}$ in the $\beta=1.0$ case and with
  $3010\,{\rm km\,s^{-1}}$ for $\beta=0.3$, while without the
  additional scalar force, the velocity reaches only $2600\,{\rm
  km\,s^{-1}}$ at a time with an equal separation between bullet and
  parent cluster.}
\label{FigPosFifth}
\end{figure}

The different timing of the merger is shown in more detail in
Figure~\ref{FigPosFifth}, where we plot the coordinates of the mass
centroids of bullet and parent cluster as a function of time,
comparing results for $\beta=1$, $\beta=0.3$ and $\beta=0$ (i.e. no
fifth force).  The times when the separations reach $720\,{\rm kpc}$
are shown with vertical dotted lines.  Clearly, the $\beta=1.0$ model
leads to a significantly earlier core passage and matching time than
the other models, as a result of the higher velocity reached by the
bullet. In fact, at the time of best-match, the dark matter of the
bullet is moving with $\sim 3800\,{\rm km\,s^{-1}}$, which is here by
chance quite close to the numerical value obtained for the shock
speed. This corresponds to a substantial increase of $\Delta v_{\rm
boost} = 1200\,{\rm km\,s^{-1}} $ relative to the value predicted for
the corresponding $\Lambda$CDM run.  If $\beta$ is lowered to
$\beta=0.3$, the velocity increase is reduced to $\Delta v_{\rm boost}
= 410\,{\rm km\,s^{-1}}$, and the opening angle of the bow shock
becomes larger (see Fig.~\ref{FigXrayMassContWithFifthForce}). The
latter is hence in principle an interesting diagnostic for the
strength of the fifth force.

The above clearly shows the substantial impact models with non-trivial
interactions in the dark sector have on cosmological structure formation, and
in particular on the non-linear regime that governs the internal structure and
dynamics of halos. So far, comparatively little simulation work has addressed
this problem \citep[with notable exceptions, e.g.][]{nusser2005,Kesden2006},
but as seen here, direct numerical simulation methods are a powerful tool to
study the implications of such models and to constrain them with observational
data, so future work in this area should be very worthwhile.

\section{Conclusions}   \label{disc}

The bullet cluster 1E0657--56 offers a unique perspective into cluster physics
and the hierarchical build-up of cosmic structure. Deep X-ray observations
have revealed a strong bow shock in front of the bullet, and a contact
discontinuity, a `cold front', that gives its bright central region a sharp,
wedge-like edge. In addition, mass reconstructions based on weak- and strong
gravitational lensing have allowed a reliable determination of the projected
mass distribution of the system. The latter features a spatial offset between
the mass centroids and the baryonic mass centroids,  which represents strong
evidence for the existence of a non-baryonic dark matter component.

This combination of powerful observations destines 1E0657--56 to serve as
laboratory for constraining fundamental aspects of cluster physics and of our
cosmological model. For example, the system has already been used to determine
constraints on the collisional cross-section of dark matter
\citep{markevitch04}. It also has become one of the hardest challenges for
alternative gravity models like MOND that try to circumvent the need for dark
matter. While claims have been made that MONDian-like theories could possibly
reproduce the bullet cluster as well \citep{Angus2006,Brownstein2007}, nothing
that comes close to the quality of the matches presented here for $\Lambda$CDM
has been presented thus far to support such scenarios.

The point we have focused on in this paper is the high inferred shock
velocity in the bullet cluster, which is a direct consequence of the
measured Mach number and the pre-shock temperature of the gas. The
literature thus far has assumed that this shock velocity can be
directly identified with the velocity of the bullet subcluster itself.
Theoretical papers have then pointed out that such a large velocity
for the bullet is uncomfortably high, either implying that the system
is a rather rare outlier \citep{hayashi06} or may need new physics in
the dark sector to be explained \citep{farrar06}.

Using explicit hydrodynamical simulation models of cluster mergers, in
3D and with self-consistent gravitational dynamics of dark matter and
baryons, we have here shown that a direct correspondence between the
inferred shock velocity of the bullet and that of its associated mass
is however not given. In the case of 1E0657--56, the two velocities
can differ both because the upstream gas is accelerated towards the
shock-front by gravitational effects that can be felt ahead of the
hydrodynamical shock, and secondly, because the shock front is moving
with a different speed than the mass centroid of the bullet
subcluster. Our simulation model based on a 1:10 merger of ordinary
$\Lambda$CDM halos on a central, zero-energy orbit reproduces the
inferred shock speed well, even though the subcluster moves only with
$2600\,{\rm km\,s^{-1}}$. At the same time, the model provides
surprisingly good matches to the observed temperature profile across
the shock and cold fronts, and to the observational determined mass
distributions in the gas from X-rays, and in the total mass from
lensing.  We conclude that the bullet cluster 1E0657--56 is consistent
with the simplest possible scenario one can construct in $\Lambda$CDM
for a 1:10 mass-ratio merger between a very rich cluster of galaxies
and a smaller sub-cluster. In fact, based on the results of
\citet{hayashi06} we estimate that the likelihood to find a most
massive dark matter substructure with a velocity of at least
$2860\,{\rm km\,s^{-1}}$ embedded in a parent system with $V_{200}=
1770\,{\rm km\,s^{-1}}$ at $z=0.296$ is as high as $f\simeq 7\%$. Here
we conservatively accounted for the increase of the relative velocity
of the subcluster relative to the parent cluster, which is bounded to
be less than $260\,{\rm km\,s^{-1}}$, depending on how much mass the
subcluster has already lost.

We have also shown that the size of the spatial offset between mass
and baryonic peak is quite sensitive to the structural details of the
merging systems. While a spatial offset between the mass centroids of
dark matter and baryons arises naturally as a result of the
hydrodynamic ram pressure experienced by the bullet's gas during core
passage, this displacement depends on the relative gas fractions and
concentrations of the merging clusters, in addition to being highly
time-variable.  This could also be the reason why some earlier
numerical models of the bullet cluster missed to detect the offset
altogether \citep{takizawa05}. As Figure~\ref{FigSeparations} shows,
for low initial concentrations of the parent cluster, the shock front
and the edge of the bullet can even overtake the mass centroid of the
cluster at late times during the merger, thereby {\em reversing the
sign} of the displacement seen by \cite{clowe06}. One is hence not
necessarily guaranteed to see a spatial offset of similar size as in
1E0657--56 between dark mass centroid and gaseous centroid in {\em
all} massive cluster mergers, and during the whole post core-passage
phase. Our results emphasize that an interpretation of highly
non-linear systems such as 1E0657--56 requires self-consistent
simulation models that account accurately {\em both} for the
gravitational and hydrodynamical dynamics of dark matter and baryons
during cluster mergers.

\acknowledgements The research of G.R.F. has been supported in part by
NSF-PHY-0401232. We thank D.~Clowe, M.~Markevitch and M.~Bradac for
useful discussions, and for providing the X-ray data shown in
Fig.~2. The simulations were performed at the Computing Center of the
Max-Planck-Society, Garching, Germany.

\bibliographystyle{apj}
\bibliography{bullet.bib}

\begin{thebibliography}{38}
\expandafter\ifx\csname natexlab\endcsname\relax\def\natexlab#1{#1}\fi

\bibitem[{{Angus} {et~al.}(2006){Angus}, {Famaey}, \& {Zhao}}]{Angus2006}
{Angus}, G.~W., {Famaey}, B., \& {Zhao}, H.~S. 2006, \mnras, 371, 138

\bibitem[{{Barrena} {et~al.}(2002){Barrena}, {Biviano}, {Ramella}, {Falco}, \&
  {Seitz}}]{barrena02}
{Barrena}, R., {Biviano}, A., {Ramella}, M., {Falco}, E.~E., \& {Seitz}, S.
  2002, \aap, 386, 816

\bibitem[{{Brada{\v c}} {et~al.}(2006){Brada{\v c}}, {Clowe}, {Gonzalez},
  {Marshall}, {Forman}, {Jones}, {Markevitch}, {Randall}, {Schrabback}, \&
  {Zaritsky}}]{bradac06}
{Brada{\v c}}, M., {Clowe}, D., {Gonzalez}, A.~H., {Marshall}, P., {Forman},
  W., {Jones}, C., {Markevitch}, M., {Randall}, S., {Schrabback}, T., \&
  {Zaritsky}, D. 2006, \apj, 652, 937

\bibitem[{{Brownstein} \& {Moffat}(2007)}]{Brownstein2007}
{Brownstein}, J.~R. \& {Moffat}, J.~W. 2007, ArXiv Astrophysics e-prints

\bibitem[{{Bullock} {et~al.}(2001){Bullock}, {Kolatt}, {Sigad}, {Somerville},
  {Kravtsov}, {Klypin}, {Primack}, \& {Dekel}}]{Bullock2001}
{Bullock}, J.~S., {Kolatt}, T.~S., {Sigad}, Y., {Somerville}, R.~S.,
  {Kravtsov}, A.~V., {Klypin}, A.~A., {Primack}, J.~R., \& {Dekel}, A. 2001,
  \mnras, 321, 559

\bibitem[{{Clowe} {et~al.}(2006){Clowe}, {Brada{\v c}}, {Gonzalez},
  {Markevitch}, {Randall}, {Jones}, \& {Zaritsky}}]{clowe06}
{Clowe}, D., {Brada{\v c}}, M., {Gonzalez}, A.~H., {Markevitch}, M., {Randall},
  S.~W., {Jones}, C., \& {Zaritsky}, D. 2006, \apjl, 648, L109

\bibitem[{{Clowe} {et~al.}(2004){Clowe}, {Gonzalez}, \& {Markevitch}}]{clowe04}
{Clowe}, D., {Gonzalez}, A., \& {Markevitch}, M. 2004, \apj, 604, 596

\bibitem[{{Damour} {et~al.}(1990){Damour}, {Gibbons}, \&
  {Gundlach}}]{damour1990}
{Damour}, T., {Gibbons}, G.~W., \& {Gundlach}, C. 1990, Physical Review
  Letters, 64, 123

\bibitem[{{Dolag} {et~al.}(2002){Dolag}, {Bartelmann}, \& {Lesch}}]{Dolag2002}
{Dolag}, K., {Bartelmann}, M., \& {Lesch}, H. 2002, \aap, 387, 383

\bibitem[{{Eke} {et~al.}(2001){Eke}, {Navarro}, \& {Steinmetz}}]{Eke2001}
{Eke}, V.~R., {Navarro}, J.~F., \& {Steinmetz}, M. 2001, \apj, 554, 114

\bibitem[{{Ensslin} {et~al.}(2006){Ensslin}, {Pfrommer}, {Springel}, \&
  {Jubelgas}}]{Ensslin2006}
{Ensslin}, T.~A., {Pfrommer}, C., {Springel}, V., \& {Jubelgas}, M. 2006, ArXiv
  Astrophysics e-prints

\bibitem[{{Ettori} \& {Fabian}(2000)}]{Ettori2000}
{Ettori}, S. \& {Fabian}, A.~C. 2000, \mnras, 317, L57

\bibitem[{{Farrar} \& {Peebles}(2004)}]{farrar04}
{Farrar}, G.~R. \& {Peebles}, P.~J.~E. 2004, \apj, 604, 1

\bibitem[{{Farrar} \& {Rosen}(2006)}]{farrar06}
{Farrar}, G.~R. \& {Rosen}, R.~A. 2006, ArXiv Astrophysics e-prints

\bibitem[{{Gubser} \& {Peebles}(2004)}]{gubser2004}
{Gubser}, S.~S. \& {Peebles}, P.~J.~E. 2004, \prd, 70, 123511

\bibitem[{{Hayashi} \& {White}(2006)}]{hayashi06}
{Hayashi}, E. \& {White}, S.~D.~M. 2006, \mnras, 370, L38

\bibitem[{{Jubelgas} {et~al.}(2004){Jubelgas}, {Springel}, \&
  {Dolag}}]{Jubelgas2004}
{Jubelgas}, M., {Springel}, V., \& {Dolag}, K. 2004, \mnras, 351, 423

\bibitem[{{Jubelgas} {et~al.}(2006){Jubelgas}, {Springel}, {Ensslin}, \&
  {Pfrommer}}]{Jubelgas2006}
{Jubelgas}, M., {Springel}, V., {Ensslin}, T.~A., \& {Pfrommer}, C. 2006, ArXiv
  Astrophysics e-prints

\bibitem[{{Kesden} \& {Kamionkowski}(2006)}]{Kesden2006}
{Kesden}, M. \& {Kamionkowski}, M. 2006, \prd, 74, 083007

\bibitem[{{Lyutikov}(2006)}]{Lyutikov2006}
{Lyutikov}, M. 2006, \mnras, 373, 73

\bibitem[{{Markevitch}(2006)}]{markevitch06}
{Markevitch}, M. 2006, in ESA SP-604: The X-ray Universe 2005, ed. A.~{Wilson},
  723

\bibitem[{{Markevitch} {et~al.}(2004){Markevitch}, {Gonzalez}, {Clowe},
  {Vikhlinin}, {Forman}, {Jones}, {Murray}, \& {Tucker}}]{markevitch04}
{Markevitch}, M., {Gonzalez}, A.~H., {Clowe}, D., {Vikhlinin}, A., {Forman},
  W., {Jones}, C., {Murray}, S., \& {Tucker}, W. 2004, \apj, 606, 819

\bibitem[{{Markevitch} {et~al.}(2002){Markevitch}, {Gonzalez}, {David},
  {Vikhlinin}, {Murray}, {Forman}, {Jones}, \& {Tucker}}]{markevitch02}
{Markevitch}, M., {Gonzalez}, A.~H., {David}, L., {Vikhlinin}, A., {Murray},
  S., {Forman}, W., {Jones}, C., \& {Tucker}, W. 2002, \apjl, 567, L27

\bibitem[{{Markevitch} \& {Vikhlinin}(2007)}]{Markevitch2007}
{Markevitch}, M. \& {Vikhlinin}, A. 2007, ArXiv Astrophysics e-prints

\bibitem[{{McCarthy} {et~al.}(2006){McCarthy}, {Bower}, \&
  {Balogh}}]{McCarthy2006}
{McCarthy}, I.~G., {Bower}, R.~G., \& {Balogh}, M.~L. 2006, ArXiv Astrophysics
  e-prints

\bibitem[{{Navarro} {et~al.}(1996){Navarro}, {Frenk}, \& {White}}]{Navarro1996}
{Navarro}, J.~F., {Frenk}, C.~S., \& {White}, S.~D.~M. 1996, \apj, 462, 563

\bibitem[{{Navarro} {et~al.}(1997){Navarro}, {Frenk}, \& {White}}]{navarro97}
---. 1997, \apj, 490, 493

\bibitem[{{Nusser} {et~al.}(2005){Nusser}, {Gubser}, \& {Peebles}}]{nusser2005}
{Nusser}, A., {Gubser}, S.~S., \& {Peebles}, P.~J. 2005, \prd, 71, 083505

\bibitem[{{Pfrommer} {et~al.}(2006){Pfrommer}, {Ensslin}, {Springel},
  {Jubelgas}, \& {Dolag}}]{Prommer2006}
{Pfrommer}, C., {Ensslin}, T.~A., {Springel}, V., {Jubelgas}, M., \& {Dolag},
  K. 2006, ArXiv Astrophysics e-prints

\bibitem[{{Sijacki} \& {Springel}(2006)}]{Sijacki2006}
{Sijacki}, D. \& {Springel}, V. 2006, \mnras, 371, 1025

\bibitem[{{Spergel} {et~al.}(2006){Spergel}, {Bean}, {Dore'}, {Nolta},
  {Bennett}, {Hinshaw}, {Jarosik}, {Komatsu}, {Page}, {Peiris}, {Verde},
  {Barnes}, {Halpern}, {Hill}, {Kogut}, {Limon}, {Meyer}, {Odegard}, {Tucker},
  {Weiland}, {Wollack}, \& {Wright}}]{spergel06}
{Spergel}, D.~N., {Bean}, R., {Dore'}, O., {Nolta}, M.~R., {Bennett}, C.~L.,
  {Hinshaw}, G., {Jarosik}, N., {Komatsu}, E., {Page}, L., {Peiris}, H.~V.,
  {Verde}, L., {Barnes}, C., {Halpern}, M., {Hill}, R.~S., {Kogut}, A.,
  {Limon}, M., {Meyer}, S.~S., {Odegard}, N., {Tucker}, G.~S., {Weiland},
  J.~L., {Wollack}, E., \& {Wright}, E.~L. 2006, ArXiv Astrophysics e-prints

\bibitem[{{Springel}(2005)}]{springel05gadget}
{Springel}, V. 2005, \mnras, 364, 1105

\bibitem[{{Springel} {et~al.}(2005{\natexlab{a}}){Springel}, {Di Matteo}, \&
  {Hernquist}}]{springel05}
{Springel}, V., {Di Matteo}, T., \& {Hernquist}, L. 2005{\natexlab{a}}, \mnras,
  361, 776

\bibitem[{{Springel} {et~al.}(2005{\natexlab{b}}){Springel}, {White},
  {Jenkins}, {Frenk}, {Yoshida}, {Gao}, {Navarro}, {Thacker}, {Croton},
  {Helly}, {Peacock}, {Cole}, {Thomas}, {Couchman}, {Evrard}, {Colberg}, \&
  {Pearce}}]{SpringelMill2005}
{Springel}, V., {White}, S.~D.~M., {Jenkins}, A., {Frenk}, C.~S., {Yoshida},
  N., {Gao}, L., {Navarro}, J., {Thacker}, R., {Croton}, D., {Helly}, J.,
  {Peacock}, J.~A., {Cole}, S., {Thomas}, P., {Couchman}, H., {Evrard}, A.,
  {Colberg}, J., \& {Pearce}, F. 2005{\natexlab{b}}, \nat, 435, 629

\bibitem[{{Springel} {et~al.}(2001){Springel}, {Yoshida}, \&
  {White}}]{springel01}
{Springel}, V., {Yoshida}, N., \& {White}, S.~D.~M. 2001, New Astronomy, 6, 79

\bibitem[{{Takizawa}(2005)}]{takizawa05}
{Takizawa}, M. 2005, \apj, 629, 791

\bibitem[{{Vikhlinin} {et~al.}(2001{\natexlab{a}}){Vikhlinin}, {Markevitch}, \&
  {Murray}}]{Vikhlinin2001}
{Vikhlinin}, A., {Markevitch}, M., \& {Murray}, S.~S. 2001{\natexlab{a}}, \apj,
  551, 160

\bibitem[{{Vikhlinin} {et~al.}(2001{\natexlab{b}}){Vikhlinin}, {Markevitch}, \&
  {Murray}}]{Vikhlinin2001b}
---. 2001{\natexlab{b}}, \apjl, 549, L47

\end{thebibliography}

\end{document}